\documentclass[10pt]{article}
\usepackage{geometry,url,graphicx}
\usepackage{amsmath,amssymb,amsthm,enumerate}
\usepackage[sort&compress]{natbib}
\usepackage{newcent}
\usepackage{color}
\def\blue#1{\textcolor{blue}{#1}}
\def\red#1{\textcolor{red}{#1}}
\definecolor{dartmouthgreen}{rgb}{0.05, 0.5, 0.06}
\definecolor{ferngreen}{rgb}{0.31, 0.47, 0.26}
\definecolor{shamrockgreen}{rgb}{0.0, 0.62, 0.38}
\definecolor{darkgreen}{rgb}{0.0, 0.2, 0.13}
\definecolor{darkspringgreen}{rgb}{0.09, 0.45, 0.27}
 \definecolor{dkg}{rgb}{0,0.7,0}

\usepackage{hyperref}
\hypersetup{colorlinks=true,allcolors=blue}

\def\today{\number\day\ \ifcase\month\or
	January\or February\or March\or April\or May\or June\or
	July\or August\or September\or October\or November\or December\fi
	\space \number\year} 
\def\gobble#1#2{}
\def\shortdate{\number\day/\number\month/\expandafter\gobble\number\year}

\usepackage{fancyhdr}

\def\rmd{\mathrm{d}}
\def\rme{\mathrm{e}}
\def\rmi{\mathrm{i}}
\def\ds{\displaystyle}

\def\PI{\mbox{\rm P$_{\rm I}$}}
\def\PII{\mbox{\rm P$_{\rm II}$}}
\def\PIII{\mbox{\rm P$_{\rm III}$}}
\def\PIV{\mbox{\rm P$_{\rm IV}$}}
\def\PV{\mbox{\rm P$_{\rm V}$}}
\def\PVI{\mbox{\rm P$_{\rm VI}$}}

\def\a{\alpha}
\def\b{\beta}
\def\ga{\gamma}
\def\de{\delta}
\def\ep{\varepsilon}
\def\ph{\varphi}
\def\k{\kappa}

\def\la{\lambda}
\def\th{\theta}

\def\p{Painlev\'{e}}

\def\pdes{partial differential equations}
\def\peq{\p\ equation}
\def\peqs{\p\ equations}

\def\bk{B\"ack\-lund}
\def\bt{B\"ack\-lund transformation}
\def\bts{B\"ack\-lund transformations}

\def\hide#1{}

\def\hf{\tfrac{1}{2}}
\def\hfz{\tfrac{1}{2}z}

\newcommand{\Integer}{\mathbb{Z}}

\newcommand{\deriv}[3][]{\frac{\rmd^{#1}{#2}}{{\rmd{#3}}^{#1}}}
\newcommand{\pderiv}[3][]{\frac{\partial^{#1}{#2}}{{\partial{#3}}^{#1}}}

\def\wz{\deriv{w}{z}}
\def\wzz{\deriv[2]{w}{z}}
\def\Lag#1#2{L_{#1}^{(#2)}}

\newtheorem{theorem}{Theorem}[section]

\newtheorem{lemma}[theorem]{Lemma}
\newtheorem{corollary}[theorem]{Corollary}
\theoremstyle{definition}
\newtheorem{definition}[theorem]{Definition}
\newtheorem{example}[theorem]{Example}
\newtheorem{remark}[theorem]{Remark}
\newtheorem{remarks}[theorem]{Remarks}

\numberwithin{figure}{section}
\numberwithin{equation}{section}
\numberwithin{table}{section}

\newcommand{\comment}[1]{}
\def\beq{\begin{equation}}
\def\eeq{\end{equation}}
\def\wh#1{\widehat{#1}}
\def\wt#1{\widetilde{#1}}
\def\pms#1#2#3{(\a,\b,\ga)=\big(#1,#2,#3\big)}%{\a=#1,\quad \b=#2,\quad \ga=#3}%,\quad D=-\hf}
\def\Lag#1#2{L_{#1}^{(#2)}}
\def\Tmn#1#2{T_{#1}^{(#2)}}
\def\Umn#1#2{U_{#1}^{(#2)}}

\def\Bmn#1#2{\mathcal{B}_{#1}^{(#2)}}

\usepackage{tikz} 
\usepackage{youngtab}

\DeclareMathOperator{\Wr}{Wr}
\DeclareMathOperator{\Top}{T}

\def\=#1{\overline{#1}}

\newcommand{\poch}[2]{(#1)_{#2}}

\def\A#1{\widetilde{A}_{#1}^{(1)}}
\def\spv{s\PV}

\newcommand{\sbr}[1]{\left(#1\right)}

\newcommand{\BesselZ}[2]{\mathcal{Z}_{#1}\!\left(#2\right)}
\newcommand{\KummerM}[3]{M\!\left(#1,#2,#3\right)}

\def\bhfz{\big(\hfz\big)}
\newcommand{\Zn}[2][]{\mathcal{Z}_{#2}^{#1}\bhfz}
\newcommand{\Zx}[2][]{\mathcal{Z}_{#2}^{#1}\big(\hf x\big)}
\def\Tr{\mathcal{T}}
\def\Nat{\mathbb{N}}
\newcommand{\Matrix}[1]{\ifcat#1a\mathbf{#1}\else\boldsymbol{#1}\fi}
\newcommand{\Vector}[1]{\ifcat#1a\mathbf{#1}\else\boldsymbol{#1}\fi}
\def\bold#1{\mbox{\boldmath $#1$}}

\begin{document}

\title{B\"{a}cklund transformations of the fifth Painlev\'{e} equation}

\author{Peter A.\ Clarkson$^{1}$\footnote{Corresponding author}, Clare Dunning$^{2}$ and Ben Mitchell$^{1}$\\[5pt]
$^{1}$School of Engineering, Mathematics and Physics,\\ University of Kent, Canterbury, CT2 7NF, UK\\[5pt] 
$^{2}$School of Mathematics, University of Leeds, Leeds, LS2 9JT, UK\\[5pt]
Email: {P.A.Clarkson@kent.ac.uk, T.C.Dunning@leeds.ac.uk, bsjm2@kent.ac.uk}%\\
}
\maketitle
 
\begin{abstract} 
In this paper B\"{a}cklund transformations of the fifth Painlev\'{e} equation are discussed.
The B\"{a}cklund transformations are applied to rational solutions of the fifth Painlev\'{e} equation, which are expressed in terms of the generalised Laguerre polynomials and generalised Umemura polynomials.
Using appropriate B\"{a}cklund transformations, we derive non-unique solutions of the fifth Painlev\'{e} equation in terms of modified Bessel functions. Further it is shown that there are also rational solutions in terms of generalised Umemura polynomials for the same values of the parameters, which are not special cases of the modified Bessel function solutions.
The relationship between solutions of the fifth Painlev\'{e} equation and solutions of the third Painlev\'{e} equation is also discussed.
\end{abstract}



\tableofcontents

\section{Introduction}
In this paper we are concerned with \bts\ associated with the fifth \peq %is given by
\beq\wzz=\left(\frac{1}{2w}+\frac{1}{w-1}\right)\!\left(\wz\right)^{\!\!2}-\frac{1}{z}\wz+\frac{(w-1)^2(\a w^2+{\b})}{z^2w}+\frac{\ga w}{z} 
+\frac{\de w(w+1)}{w-1},\label{eq:p5gen}\eeq
with $\a$, $\b$, $\ga$ and $\de$ constants. Primarily we study the generic case of \eqref{eq:p5gen} when $\de\not=0$, so we set
$\de=-\hf$, without loss of generality (by rescaling $z$ if necessary) and obtain
\beq\wzz=\left(\frac{1}{2w}+\frac{1}{w-1}\right)\!\left(\wz\right)^{\!\!2}-\frac{1}{z}\wz+\frac{(w-1)^2(\a w^2+{\b})}{z^2w}+\frac{\ga w}{z}-\frac{w(w+1)}{2(w-1)},\label{eq:p5}\eeq
which henceforth we refer to as \PV. The case when $\de=0$ in \eqref{eq:p5gen}, %setting $\a=A$, $\b=B$ and $\ga=C$,
\beq\wzz=\left(\frac{1}{2w}+\frac{1}{w-1}\right)\!\left(\wz\right)^{\!\!2}-\frac{1}{z}\wz+\frac{(w-1)^2(\a w^2+\b)}{z^2w}+\frac{\ga w}{z},%\frac{(w-1)^2(A w^2+B)}{z^2w}+\frac{C w}{z},
\label{eq:p5deg}\eeq
is known as the \textit{degenerate fifth \peq}\ (deg-\PV) and is related to the third \peq\ (\PIII)
\beq \deriv[2]{q}{x}=\frac{1}{q}\!\left(\deriv{q}{x}\right)^{\!2}-\frac{1}{x}\deriv{q}{x}+\frac{A q^2+B}{x}+C q^3+\frac{D}{q},\label{eq:p3} \eeq
where $A$, $B$, $C$ and $D$ are constants.
If $\ga=\de=0$ in \eqref{eq:p5gen}, then it has the first integral
\[ z^2\left(\wz\right)^{\!2}=2(w-1)^2(\a w^2+\k w-\b),\]
with $\k$ a constant of integration, which is solvable in terms of elementary functions.

An important application of \bts\ of \peqs\ is to generate hierarchies of solutions. Using a ``seed solution", typically a simple solution of the \peq, the successive applications of \bts\ generates a hierarchy of solutions which frequently can be expressed as a determinant, cf.~\cite{refPAC06review,refGLS}.
A second application of \bts\ of \peqs\ is to derive discrete \peqs. For \PV\ \eqref{eq:p5} associated discrete equations are derived in \cite{refCDM26,refFGR,refTGR02}.
\bts\ also provide a structure for the parameter space of \peqs. Okamoto 
\cite{refOkamotoPV} showed that the \bts\ of \PV\ \eqref{eq:p5} form the extended affine Weyl group $\widetilde{W}\big(A_3^{(1)}\big)$, which is discussed further in \S\ref{ssec:spv}. 

The fifth \peq\ \eqref{eq:p5gen} arises in numerous applications including: 
general relativity \cite{refLM79,refLM82,refLM83,refLM84,refLM85,refLM86,refLM87,refMD79,refMarek}; 
number theory \cite{refBGR18}; 
orthogonal polynomials \cite{refBC09,refBC15,refBCE10,refBFSVAZ,refBFVA,refBVA,refCCF,refChenDai,refPAC13,refPAC23,refFIM25,refFMMR,refFvA11,refFvA13,refWVAbk,refXZ13,refXZ15}; 
random matrices \cite{refACRS,refABKSW,refAGKW26a,refAGKW26b,refBBBCPRS,refBBBGIIK,refBCZ,refBothWa,refBothWb,refCF,refCIK,refCK,refForrBook,refFMay,refFW02,refFW04,refFW06,refFW07,refTW94b,refWFC00}; 
statistical physics \cite{refAresViti,refEFIK,refJMMS,refKanz}; 
supersymmetric quantum mechanics \cite{refBF,refBFN,refFMS16,refMarq,refM-S}; and
vortex solutions of complex sine-Gordon equations \cite{refBP98,refPAC23,refOB05}.
We remark that several of these applications involve the \PV\ $\sigma$-equation. It is the second-order, second-degree differential equation derived in \cite{refJMi,refOkamoto80a,refOkamoto80b,refOkamotoPV} which is associated with the Hamiltonian representation of \PV\ given by
\beq\label{eq:S5JM}
\left(z\deriv[2]{\sigma}{z}\right)^{\!\!2}= \left[ 2\left(\deriv{\sigma}{z}\right)^{\!\!2} 
%+(\k_1+\k_2+\k_3-z)\deriv{\sigma}{z}+\sigma\right]^2
-z\deriv{\sigma}{z}+\sigma\right]^2
-4\prod_{j=0}^3\left(\deriv{\sigma}{z}+\k_{j}\right),\eeq
with $\k_0$, $\k_1$, $\k_2$ and $\k_3$ constants such that $\k_0+\k_1+\k_2+\k_3=0$, 
whose solution is expressible in terms of the solution of \PV\ \eqref{eq:p5}.

The fifth \peq\ \eqref{eq:p5gen} also arises as a symmetry reduction of various nonlinear \pdes\ including:
the Einstein-Weyl equation \cite{refSchief00,refSchief02,refTTX};
the Maxwell-Bloch system with pumping \cite{refBurtsev};
the three-wave resonant interaction system \cite{refJKT,refMW}; 
the modified Yajima-Oikawa equation \cite{refKIK}; 
shallow water wave equations \cite{refCMsww,refCMacta};
the $2+1$-dimensional Sine-Gordon equation \cite{refCMM}; and
stimulated-Raman-scattering equations \cite{refLMW}.

The outline of this paper is as follows.
In \S\ref{sec:btst} we describe \bk\ and Schlesinger transformations of \PV\ \eqref{eq:p5}. 
In \S\ref{sec:rats} we apply \bk\ and Schlesinger transformations to rational solutions of \PV\ \eqref{eq:p5}. 
In \S\ref{sec:Bessel_non-unique} we show that for certain values of the parameters there are non-unique solutions of \PV\ \eqref{eq:p5} which are expressed in terms of modified Bessel functions. Further we show that for these parameters there also is a rational solution which is \textit{not} a special case of either modified Bessel function solution.
In \S\ref{sec:PVPIII} we discuss the relationship between solutions of the third and fifth \peqs. 
In \S\ref{sec:dis} we discuss our results.

\section{\label{sec:btst}\bk\ and Schlesinger transformations}
\subsection{\label{ssec:bt}\bts}
In this section we describe the \bts\ of \PV\ \eqref{eq:p5}. 
\begin{definition} A \textit{\bt} maps a solution of a given \peq\ to another solution of the same \peq, with different values of the parameters, or another equation.
\end{definition}

First we note that solutions of \PV\ \eqref{eq:p5} have the symmetries 
\begin{subequations}\label{PV:Sym}\begin{align}
\label{PV:Sym1}\mathcal{S}_{1}:\qquad &{w_{1}}({z};{\a_{1}}, {\b_{1}}, {\ga_{1}})
={w(-z;\a,\b,\ga)},&&({\a_{1}}, {\b_{1}}, {\ga_{1}})=(\a,\b,-\ga),\\
\label{PV:Sym2}\mathcal{S}_{2}:\qquad &{w_{2}}({z};{\a_{2}}, {\b_{2}}, {\ga_{2}})
=\frac{1}{w(z;\a,\b,\ga)},&&({\a_{2}}, {\b_{2}}, {\ga_{2}})=(-\b,-\a,-\ga),
\end{align}\end{subequations}
where $w(z;\a,\b,\ga)$ is a solution of \PV\ \eqref{eq:p5}. If two solutions of \PV\ are related through these symmetries then we regard them as equivalent as essentially they are the same solution.

%\begin{theorem}\label{thm:bt2}{
\begin{definition}\label{def:btpv}
Let $w=w(z;\a,\b,\ga)$ be a solution of \PV\ \eqref{eq:p5} such that
\[%\beq\label{eq:PT.BTs.P5a}
\Phi(w;z)=z\wz-\ep_{2}aw^2+(\ep_{2}a-\ep_{3}b+\ep_{1}z)w+\ep_{3}b \not=0,\]%\eeq
and \[%\beq\label{eq:PT.BTs.P5a1}
\Phi(w;z)-2\ep_{1}z w\not=0,\]%\eeq
where $a=\sqrt{2\a}$ and $b=\sqrt{-2\b}$ and $c=\ga$. Then the \textit{\bt}\ $\Tr_{\ep_{1},\ep_{2},\ep_{3}}$ is defined by
\begin{subequations}\label{eq:PT.BTs.P5}\beq\label{eq:PT.BTs.P5b}
\Tr_{\ep_{1},\ep_{2},\ep_{3}}:\quad
W(z;A,B,C)%=\Tr_{\ep_{1},\ep_{2},\ep_{3}}(w)
=1-\frac{2\ep_{1}z w}{\Phi(w;z)}
=1-\frac{2\ep_{1}z w}{z w'-\ep_{2}aw^2+(\ep_{2}a-\ep_{3}b+\ep_{1}z)w+\ep_{3}b},
\eeq
with $'\equiv\rmd/\rmd z$, $\ep_j=\pm1$, $j=1,2,3$, independently, and for the parameters
\beq\label{eq:PT.BTs.P5c}
\Tr_{\ep_{1},\ep_{2},\ep_{3}}(a,b,c)=\Big(\hf\big[c+\ep_{1}(1-\ep_{3}b-\ep_{2}a)\big],\hf\big[c-\ep_{1}(1-\ep_{3}b-\ep_{2}a)\big],\ep_{1}(\ep_{3}b-\ep_{2}a)\Big),\eeq
\end{subequations}
so
\[(A,B,C)=\Big(\tfrac{1}{8}\big[c+\ep_{1}(1-\ep_{3}b-\ep_{2}a)\big]^2,-\tfrac{1}{8}\big[c-\ep_{1}(1-\ep_{3}b-\ep_{2}a)\big]^2,\ep_{1}(\ep_{3}b-\ep_{2}a)\Big),\]
cf.~\cite{refGromak76,refGromak,refGF01a,refGF01b}, \cite[\S39]{refGLS} and \cite[\S32.7(v)]{refDLMF}.
\end{definition}

Note that we are using the same notation as in \cite[\S32.7(v)]{refDLMF} which slightly differs from that used in \cite{refGromak,refGF01a,refGF01b,refGLS}. %or \cite[\S39]{refGLS}.

\begin{remarks}\textrm{
\begin{enumerate}
\item Since $\ep_j=\pm1$, $j=1,2,3$, independently, there are eight distinct transformations $\Tr_{\ep_{1},\ep_{2},\ep_{3}}$ \eqref{eq:PT.BTs.P5}. The parameter $\ep_{1}$ reflects that the transformation $(z,\ga)\to(-z,-\ga)$ is a discrete symmetry of \PV\ \eqref{eq:p5}. The parameters $\ep_{2}$ and $\ep_{3}$ reflect that $\a=\hf a^2$ and $\b=-\hf b^2$, respectively, so $a$ and $b$ can take either sign.
\item The \bt\ $\Tr_{\ep_{1},\ep_{2},\ep_{3}}$ is usually given for the parameters $\a$, $\b$ and $\ga$, in terms of $a$, $b$ and $c$. However in order to apply a sequence of \bts, it is preferable to define the action of \bt\ on the parameters $a$, $b$ and $c$, rather than $\a$, $\b$ and $\ga$, to avoid any ambiguity in taking a square root.
\item The discrete symmetries of \PV\ \eqref{eq:p5}, corresponding to the extended affine Weyl group of type $A_{3}^{(1)}$, act naturally on the root variables, see, for example, \cite{refCDHM,refKNY}. Hence it is natural to define the \bts\ for \PV\ \eqref{eq:p5} in terms of the parameters $a=\sqrt{2\a}$, $b=\sqrt{-2\b}$ and $c=\ga$, rather than $\a$, $\b$ and $\ga$.
\comment{\item We note that the Hamiltonian for \PV\ \eqref{eq:p5} in \cite{refJMi} is written in terms of the monodromy parameters $\th_{0}$, $\th_{1}$ and $\th_{\infty}$, which are related to the parameters $\a$, $\b$ and $\ga$ through
\beq \a=\tfrac{1}{8}(\th_{0}-\th_{1}+\th_{\infty})^2,\qquad \b=-\tfrac{1}{8}(\th_{0}-\th_{1}-\th_{\infty})^2,\qquad\ga=1-\th_{0}-\th_{1},\label{SchlesParams}\eeq
see also \cite{refOkamotoPV}, as are the Schlesinger transformations, which relate two solutions of the associated monodromy problem, for \PV\ \eqref{eq:p5}, see \cite{refFMA,refMF}. }
\item When applying the \bt\ $\Tr_{\ep_{1},\ep_{2},\ep_{3}}$ followed by the \bt\ $\Tr_{\sigma_{1},\sigma_{2},\sigma_{3}}$ we use the notation $\Tr_{\sigma_{1},\sigma_{2},\sigma_{3}}\circ \Tr_{\ep_{1},\ep_{2},\ep_{3}}$, i.e.
\[\Tr_{\sigma_{1},\sigma_{2},\sigma_{3}}\big(\Tr_{\ep_{1},\ep_{2},\ep_{3}}(w;a,b,c)\big)=\Tr_{\sigma_{1},\sigma_{2},\sigma_{3}}\circ \Tr_{\ep_{1},\ep_{2},\ep_{3}}(w;a,b,c).\]
Further if 
\[ \Tr_{\ep_{1},\ep_{2},\ep_{3}}(w;a,b,c)= \Big(W(w,w',a,b,c);A(a,b,c),B(a,b,c),C(a,b,c)\Big),\]
then
\[\Tr_{\sigma_{1},\sigma_{2},\sigma_{3}}\circ \Tr_{\ep_{1},\ep_{2},\ep_{3}}(w;a,b,c)= \Tr_{\sigma_{1},\sigma_{2},\sigma_{3}}\Big(W(w,w',a,b,c);A(a,b,c),B(a,b,c),C(a,b,c)\Big).\]
\item Gordoa, Joshi and Pickering \cite{refGJP01a,refGJP01b} use a different approach which is a truncation-type method to derive \bts\ of \PV\ \eqref{eq:p5}.
\end{enumerate}
}\end{remarks}

\begin{example} Suppose that $w(z)$ is a solution of \PV\ \eqref{eq:p5} with $\pms{\hf a^2}{-\hf b^2}{c}$ then applying the \bt\ $\Tr_{1,1,1}$ several times gives
\begin{align*}
\Tr_{1,1,1}(w;a,b,c)&=\frac{\ds zw' - aw^2 +(a-b -z)w+ b}{\ds zw' - aw^2 +(a-b +z)w+ b} %\\ &
=1-\frac{2zw}{\ds zw' - aw^2 +(a-b +z)w+ b},\\
\Tr_{1,1,1}^{\,2}(w;a,b,c)&=\frac{zw'+(b+c-1)w^2-(2b+c-1+z)w+b}{w\big[zw'-aw^2+(2a+c-1-z)w-a-c+1\big]}\\
&= \frac{1}{w}+ \frac{(a+b+c-1)(w-1)^2}{w\big[\ds zw'-aw^2+(2a+c-1-z)w-a-c+1\big]} ,\\
\Tr_{1,1,1}^{\,3}(w;a,b,c)%&=\frac{z^2\big(w'\big)^{\!2} -2bz(w-1)w'-z^2w^2+2(b+c-1)z(w-1)w^2-(w-1)^2(a^2w^2-b^2)}{\big[zw'-aw^2+(a-b-z)w+b\big]\big[zw'+aw^2-(a+b+z)w+b\big]} \\
&= 1 + \frac{1}{a}\left\{\frac{(a+b+c-1)zw}{\ds zw'-aw^2+(a-b-z)w+b}+ \frac{(a-b-c+1)zw}{\ds zw'+aw^2-(a+b+z)w+b}\right\},
\end{align*}
with $'=\rmd/\rmd z$, provided that the denominators do not vanish, and the parameters $(a,b,c)$ transform as 
\begin{align*}
\Tr_{1,1,1}(a,b,c)&=\big(\hf(c-a-b+1),\hf(a+b+c-1),b-a\big),\\
\Tr_{1,1,1}^{\,2}(a,b,c)&=\big(\hf(b-a-c+1),\hf(b-a+c-1),a+b-1\big),\\
\Tr_{1,1,1}^{\,3}(a,b,c)&=\big(a,b-1,c-1\big).
\end{align*}
We note that $\Tr_{1,1,1}^{\,2}(w)$ is not quadratic in %$\rmd w/\rmd z$ 
$w'$ as might have been expected.
\end{example}

\begin{lemma}If $\Tr_{\ep_{1},\ep_{2},\ep_{3}}$ is given by \eqref{eq:PT.BTs.P5} and $\mathcal{S}_{2}$ is the symmetry \eqref{PV:Sym2}, then
\begin{align*} 
\Tr_{-1,1,1}=\mathcal{S}_{2}\circ\Tr_{1,1,1},\quad \Tr_{-1,-1,1}=\mathcal{S}_{2}\circ\Tr_{1,-1,1},\quad
\Tr_{1,1,-1}=\mathcal{S}_{2}\circ\Tr_{-1,1,-1}, \quad \Tr_{1,-1,-1}=\mathcal{S}_{2}\circ\Tr_{-1,-1,-1}.
\end{align*}
\end{lemma}

\begin{proof} Suppose that $w(z)$ is a solution of \PV\ \eqref{eq:p5} with $\pms{\hf a^2}{-\hf b^2}{c}$ then
\begin{align*} 
\Tr_{1,1,1}(w;a,b,c) &= \left(\frac{\ds zw' - aw^2 +(a-b -z)w+ b}{\ds zw' - aw^2 +(a-b +z)w+ b};\hf(c-a-b+1),\hf(a+b+c-1),b-a \right),\\
\Tr_{-1,1,1}(w;a,b,c) &= \left(\frac{\ds zw' - aw^2 +(a-b +z)w+ b}{\ds zw' - aw^2 +(a-b -z)w+ b};\hf(a+b+c-1),\hf(c-a-b+1),a-b\right).
\end{align*}
If we define \[ \mathcal{S}_{2}(w;a,b,c) = (1/w;b,a,-c),\] then applying $\mathcal{S}_{2}$ to $ \Tr_{1,1,1}(w;a,b,c)$ shows that $\Tr_{-1,1,1}=\mathcal{S}_{2}\circ\Tr_{1,1,1}$. The other identities are proved in a similar way.
\end{proof}

To generate a hierarchy of rational solutions, or special function solutions, one needs to change at least one of the parameters $a$, $b$ and $c$ by $\pm1$.
In order to do this, it is necessary to use a combination of \bts. In Tables \ref{btab}, \ref{btac} and \ref{btbc}, the transformations that change two of $a$, $b$ and $c$ by $\pm1$ are given.

%It can be shown that applying a sequence of three \bts\ gives the following
\begin{table}[ht]
\[\begin{array}{|c|cc|ccc|} \hline 
%& a_{1} & b_{1}& \a_{1}=\tfrac12a_{1}^2 & \b_{1}=-\tfrac12b_{1}^2 & \ga_{1} \\ \hline
& a & b& \a & \b & \ga \\ \hline
\Tr_{1,-1,1}\circ\Tr_{1,1,1}(a,b,c) & -a+1 & b-1 & \tfrac12(a-1)^2 & -\tfrac12(b-1)^2 & \phantom{\Big(}c\phantom{\Big(}\\
\Tr_{-1,1,-1}\circ\Tr_{-1,1,1}(a,b,c) & a-1 & -b+1 & \tfrac12(a-1)^2 & -\tfrac12(b-1)^2 & c\\ \hline
\Tr_{1,-1,1}\circ\Tr_{1,-1,1}(a,b,c) & a+1 & b-1 & \tfrac12(a+1)^2 & -\tfrac12(b-1)^2 & c\\
\Tr_{-1,1,-1}\circ\Tr_{-1,-1,1}(a,b,c) & -a-1 & -b+1 & \tfrac12(a+1)^2 & -\tfrac12(b-1)^2 & c\\ \hline
\Tr_{1,-1,1}\circ\Tr_{1,1,-1}(a,b,c) & -a+1 & -b-1 & \tfrac12(a-1)^2 & -\tfrac12(b+1)^2 & c\\
\Tr_{-1,1,-1}\circ\Tr_{-1,1,-1}(a,b,c) & a-1 & b+1 & \tfrac12(a-1)^2 & -\tfrac12(b+1)^2 & c\\ \hline
\Tr_{1,-1,1}\circ\Tr_{1,-1,-1}(a,b,c) & a+1 & -b-1 & \tfrac12(a+1)^2 & -\tfrac12(b+1)^2 & c\\
\Tr_{-1,1,-1}\circ\Tr_{-1,-1,-1}(a,b,c) & -a-1 & b+1 & \tfrac12(a+1)^2 & -\tfrac12(b+1)^2 & c\\ 
\hline \end{array}\]
\caption{\label{btab}\bts\ that change the parameters $a$ and $b$ by $\pm1$.}
\end{table}
\begin{table}[ht]
\[\begin{array}{|c|cc|ccc|} \hline 
%& a_{1} & b_{1}& \a_{1}=\tfrac12a_{1}^2 & \b_{1}=-\tfrac12b_{1}^2 & \ga_{1} \\ \hline
& a & b& \a & \b & \ga \\ \hline
\Tr_{-1,1,1}^{\,3}(a,b,c) & a-1 & b & \hf(a-1)^2 & -\hf b^2 & c-1\\
\Tr_{1,1,1}^{\,2}\circ\Tr_{-1,1,1}(a,b,c) & -a+1 & -b & \hf(a-1)^2 & -\hf b^2 & c-1\\
\Tr_{1,1,1}^{\,2}\circ\Tr_{-1,1,-1}(a,b,c) & -a+1 & b & \hf(a-1)^2 & -\hf b^2 & c-1\\ \hline
\Tr_{1,1,1}^{\,2}\circ\Tr_{-1,-1,1}(a,b,c) & a+1 & -b & \hf(a+1)^2 & -\hf b^2 & c-1\\
\Tr_{1,1,1}^{\,2}\circ\Tr_{-1,-1,-1}(a,b,c) & a+1 & b & \hf(a+1)^2 & -\hf b^2 & c-1\\ \hline
\Tr_{-1,-1,-1}^{\,2}\circ\Tr_{1,1,1}(a,b,c) & a-1 & b & \hf(a-1)^2 & -\hf b^2 & \phantom{\Big(}c+1\phantom{\Big(}\\
\Tr_{-1,-1,-1}^{\,2}\circ\Tr_{1,1,-1}(a,b,c) & a-1 & -b & \hf(a-1)^2 & -\hf b^2 & c+1\\ \hline
\Tr_{1,-1,-1}^{\,3}(a,b,c) & a+1 & b & \hf(a+1)^2 & -\hf b^2 & c+1\\
\Tr_{-1,-1,-1}^{\,2}\circ\Tr_{1,-1,1}(a,b,c) & -a-1 & b & \hf(a+1)^2 & -\hf b^2 & c+1\\
\Tr_{-1,-1,-1}^{\,2}\circ\Tr_{1,-1,-1}(a,b,c) & -a-1 & -b & \hf(a+1)^2 & -\hf b^2 & c+1\\ 
\hline \end{array}\]
\caption{\label{btac}\bts\ that change the parameters $a$ and $c$ by $\pm1$.}
\end{table}
\begin{table}[ht]
\[\begin{array}{|c|cc|ccc|} \hline 
%& a_{1} & b_{1}& \a_{1}=\tfrac12a_{1}^2 & \b_{1}=-\tfrac12b_{1}^2 & \ga_{1} \\ \hline
& a & b& \a & \b & \ga \\ \hline
\Tr_{1,1,1}^{\,3}(a,b,c) & a & b-1 & \hf a^2 & -\hf(b-1)^2 & \phantom{\Big(}c-1\phantom{\Big(}\\
%\Tr_{1,1,1}^{\,2}\circ\Tr_{1,1,1} & a & b-1 & \hf a^2 & -\hf(b-1)^2 & \phantom{\Big(}c-1\phantom{\Big(}\\
\Tr_{1,1,1}^{\,2}\circ\Tr_{1,-1,1}(a,b,c) & -a & b-1 & \hf a^2 & -\hf(b-1)^2 & c-1\\ \hline
\Tr_{-1,1,1}^{\,2}\circ\Tr_{1,-1,-1}(a,b,c) & a & b+1 & \hf a^2 & -\hf(b+1)^2 & c-1\\
\Tr_{1,1,1}^{\,2}\circ\Tr_{1,1,-1}(a,b,c) & a & -b-1 & \hf a^2 & -\hf(b+1)^2 & c-1\\
\Tr_{1,1,1}^{\,2}\circ\Tr_{1,-1,-1}(a,b,c) & -a & -b-1 & \hf a^2 & -\hf(b+1)^2 & c-1\\ \hline
\Tr_{1,-1,-1}^{\,2}\circ\Tr_{-1,1,1}(a,b,c) & a & b-1 & \hf a^2 & -\hf(b-1)^2 & c+1\\
\Tr_{-1,-1,-1}^{\,2}\circ\Tr_{-1,1,1}(a,b,c) & -a & -b+1 & \hf a^2 & -\hf(b-1)^2 & c+1\\
\Tr_{-1,-1,-1}^{\,2}\circ\Tr_{-1,-1,1}(a,b,c) & a & -b+1 & \hf a^2 & -\hf(b-1)^2 & c+1\\ \hline
\Tr_{-1,-1,-1}^{\,2}\circ\Tr_{-1,1,-1}(a,b,c) & -a & b+1 & \hf a^2 & -\hf(b+1)^2 & c+1\\
%\Tr_{-1,-1,-1}^{\,2}\circ\Tr_{-1,-1,-1} & a & b+1 & \hf a^2 & -\hf(b+1)^2 & c+1\\ 
\Tr_{-1,-1,-1}^{\,3}(a,b,c) & a & b+1 & \hf a^2 & -\hf(b+1)^2 & c+1\\ 
\hline\end{array}\]
\caption{\label{btbc}\bts\ that change the parameters $b$ and $c$ by $\pm1$.}
\end{table}

\newpage
\begin{remarks}\textrm{\label{rem:2.3}
\begin{enumerate}%\rmd w/\rmd z
\item The \bts\ given in Tables \ref{btab}, \ref{btac} and \ref{btbc} are quadratic in $\rmd w/\rmd z$, as illustrated by the first transformation in each table. Suppose that $w$ is a solution of \PV\ \eqref{eq:p5} with $\pms{\hf a^2}{-\hf b^2}{c}$, then
\begin{align*}
\Tr_{1,-1,1}\circ\Tr_{1,1,1}(w)&=\frac{\left\{zw'+(w-1)(bw-w-b)\right\}^{2} -z^2w^2 +2czw^2(w-1)-(a+b-1)^2w^2(w-1)^2 }{w\left\{ %
\left[zw'-(w-1)(aw-a+1)\right]^{2} -z^2w^2 +2czw(w-1) -(a+b-1)^2(w-1)^2\right\}},\\
\Tr_{-1,1,1}^{\,3}(w)&=\frac{\left\{zw'-aw^2+(a+b-z)w-b\right\}\left\{zw'-aw^2+(a-b-z)w+b\right\}}{\left\{zw'-aw(w-1)\right\}^{2}-z^2w^2+2(a+c-1)zw(w-1)-b^2(w-1)^2}\\
&=1-\frac{2zw\big\{zw'-aw^2+(2a+c-z-1)w-a-c+1\big\}}{\left\{zw'-aw(w-1)\right\}^{2}-z^2w^2+2(a+c-1)zw(w-1)-b^2(w-1)^2},\\
\Tr_{1,1,1}^{\,3}(w)%=1+\frac{(a+b+c-1)zw}{a\left\{zw'-aw^2+(a-b-z)w+b\right\}} +\frac{(a-b-c+1)zw}{a\left\{zw'+aw^2-(a+b+z)w+b \right\}}, 
&=1+\frac{1}{a} \left\{\frac{(a+b+c-1)zw}{zw'-aw^2+(a-b-z)w+b} +\frac{(a-b-c+1)zw}{zw'+aw^2-(a+b+z)w+b}\right\}, 
\end{align*}
where $'\equiv\rmd/\rmd z$. 
It might have been expected that the \bts\ in Tables \ref{btac} and \ref{btbc} would be cubic in $w'$, %$\rmd w/\rmd z$, 
though frequently there are simplifications.
 In Table \ref{btab}, first apply the \bt\ $\Tr_{\ep_{1},\ep_{2},\ep_{3}}$, then apply either $\Tr_{1,-1,1}$ or $\Tr_{-1,1,-1}$. 
 In Tables \ref{btac} and \ref{btbc}, first apply the \bt\ $\Tr_{\ep_{1},\ep_{2},\ep_{3}}$, then apply one of $\Tr_{1,1,1}$, $\Tr_{-1,1,1}$, $\Tr_{1,-1,-1}$ or $\Tr_{-1,-1,-1}$ twice. 
\item There are other combinations of \bts\ which will give the same effect on the parameters. %; there are already two possible ways of achieving each result.
\item\label{rem:2.3.5} From Tables \ref{btab}, \ref{btac} and \ref{btbc}, in general two of $a$, $b$ and $c$ will change by $\pm1$. If either $a=\pm\hf$ or $b=\pm\hf$, then there exist \bts\ such that only \textit{one} of $\a$, $\b$ and $\ga$ will change. Further, if $a=\pm\hf$ and $b=\pm\hf$ then there exist \bts\ such that \textit{none} of $\a$, $\b$ and $\ga$ will change. This is illustrated in \S\ref{sec:Bessel_non-unique}, see Example \ref{example:w0}. %\red{Cross reference.}
\end{enumerate}
}\end{remarks}

\begin{example}
Suppose that $w_0=w(z;\a,-\a,0)$ is a solution of \PV\ \eqref{eq:p5} with $\pms{\hf a^2}{-\hf a^2}{0}$, then
\begin{align*}
w_{1}&=\Tr_{1,1,1}(w_0;a,a,0)=\frac{1}{\Tr_{-1,1,1}(w_0;a,a,0)}=\frac{zw'-aw^2-zw+a}{zw' -aw^2+zw+a},\\
w_{2}&=\Tr_{-1,-1,-1}(w_0;a,a,0)=\frac{1}{\Tr_{1,-1,-1}(w_0;a,a,0)}=\frac{zw'+aw^2+zw-a}{zw' +aw^2-zw-a}.
\end{align*}
The solution $w_{1}$ satisfies \PV\ \eqref{eq:p5} with $\pms{\hf (a-\hf)^2}{-\hf (a-\hf)^2}{0}$ whilst the solution $w_{2}$ satisfies \PV\ with $\pms{\hf (a+\hf)^2}{-\hf (a+\hf)^2}{0}$.
\end{example}

\subsection{\label{ssec:sch}Schlesinger transformations}
A special class of \bts\ for the \peqs\ is generated by \textit{Schlesinger transformations}, which relate two solutions of the associated monodromy problem.

Each of the \peqs\ can be expressed as the compatibility condition of a linear system, often known as the isomonodromy problem or Lax pair,
\beq\label{eq:PT.IP.gen}
\pderiv{\bold{\Psi}}{\lambda} = \Matrix{A}_\mathrm{J}(z,\lambda)\bold{\Psi},\qquad
\pderiv{\bold{\Psi}}{z} = \Matrix{B}_\mathrm{J}(z,\lambda)\bold{\Psi},\qquad \mathrm{J}=\mathrm{I, II, \ldots, VI},\eeq
where $\Matrix{A}_\mathrm{J}$ and $\Matrix{B}_\mathrm{J}$ are matrices whose
entries depend on the solution $w(z)$ of the \peq. The compatibility condition of \eqref{eq:PT.IP.gen} is given by 
\[\frac{\partial^2\bold{\Psi}}{\partial z\partial\la} =\frac{\partial^2\bold{\Psi}}{\partial\la\partial z},\]
which is satisfied provided that 
\beq\label{eq:PT.IP.comcond}
\pderiv{\Matrix{A}_\mathrm{J}}{z}-\pderiv{\Matrix{B}_\mathrm{J}}{\la}
+\Matrix{A}_\mathrm{J}\Matrix{B}_\mathrm{J}-\Matrix{B}_\mathrm{J}\Matrix{A}_\mathrm{J}=0. \end{equation} 
Matrices $\Matrix{A}_\mathrm{J}$ and $\Matrix{B}_\mathrm{J}$ satisfying \eqref{eq:PT.IP.comcond} for \PI--\PVI\ are given by Jimbo and Miwa \cite{refJMi}, see also \cite{refFIKN}, though these are not unique.
A Schlesinger transformation $\mathcal{R}$ relates two solutions $\bold{\Psi}$ and $\bold{\Psi}_{1}$ of \eqref{eq:PT.IP.gen} in the form
\[\bold{\Psi}_{1}=\mathcal{R}\bold{\Psi}.\]

For \PV\ \eqref{eq:p5} the matrices $\Matrix{A}_\mathrm{V}$ and $\Matrix{B}_\mathrm{V}$ are 
%\begin{subequations}
\begin{align*}
\Matrix{A}_\mathrm{V}(z,\lambda) &=\left(\begin{matrix}\hf z & 0\\ 0 & -\hf z\end{matrix}\right)
+\frac{1}{\lambda}\left(\begin{matrix} v+\hf\th_{0} & -u(v+\th_{0})\\ {v}/{u} & -v-\hf\th_{0} \end{matrix}\right) \nonumber\\ &\qquad 
+ \frac{1}{\lambda-1}\left(\begin{matrix}-v-\hf(\th_{0}+\th_{\infty})& uw[v+\hf(\th_{0}-\th_{1}+\th_{\infty})]\\ 
-\dfrac{v+\hf(\th_{0}+\th_{1}+\th_{\infty})}{uw} & v+\hf(\th_{0}+\th_{\infty}) \end{matrix}\right),\\
\Matrix{B}_\mathrm{V}(z,\lambda) &=\left(\begin{matrix}\hf\la & 0\\ 0 & -\hf\la\end{matrix}\right)+ 
\frac{1}{z}\left(\begin{matrix}0 & -u\big\{v+\th_{0}-w[v+\hf(\th_{0}-\th_{1}+\th_{\infty})]\big\}\\ \dfrac{v}{u}-\dfrac{v+\hf(\th_{0}+\th_{1}+\th_{\infty})}{uw}& 0 \end{matrix}\right),
\end{align*}%\end{subequations}
with $\th_{0}$, $\th_{1}$ and $\th_{\infty}$ parameters, the \textit{monodromy parameters},
then from \eqref{eq:PT.IP.gen}
\begin{subequations}\begin{align} 
z\deriv{w}{z}&=zw-2v(w-1)^2-\hf(w-1)\left[(\th_{0}-\th_{1}+\th_{\infty})w-(3\th_{0}+\th_{1}+\th_{\infty})\right],\\
z\deriv{v}{z}&=wv\left[v+\hf(\th_{0}-\th_{1}+\th_{\infty})\right]-\frac{(v+\th_{0})\left[v+\hf(\th_{0}+\th_{1}+\th_{\infty})\right]}{w},\\ 
z\deriv{u}{z}&=u\left\{-2v-\th_{0}+w\left[v+\hf(\th_{0}-\th_{1}+\th_{\infty})\right]+\frac{v+\hf(\th_{0}+\th_{1}+\th_{\infty})}{w}\right\},
\end{align}\end{subequations}
see \cite[equations C.38, C.39]{refJMi}; also \cite[Chapter 5, \S4]{refFIKN}.
%\[\mathbf{B}_{\mathrm V} (z,\lambda)=\tfrac12\lambda\,\sigma_3+\frac1z\begin{pmatrix}0 & -u\left\{v+\theta_0-w\left[v+\tfrac12(\theta_0-\theta_1+\theta_\infty)\right]\right\}\\[4pt] \dfrac{v}{u}-\dfrac{v+\tfrac12(\theta_0+\theta_1+\theta_\infty)}{uw} & 0\end{pmatrix}\]
Consequently, eliminating $v$ shows that $w$ satisfies \PV\ \eqref{eq:p5} with
\beq\pms{\tfrac{1}{8}(\th_{0}-\th_{1}+\th_{\infty})^2}{-\tfrac{1}{8}(\th_{0}-\th_{1}-\th_{\infty})^2}{1-\th_{0}-\th_{1}}.\label{SchlesParams}\eeq
%Henceforth we shall refer to $\{\th_{0},\th_{1},\th_{\infty}\}$ as the \textit{monodromy parameters} for \PV.
The main characteristic of a Schlesinger transformation for \PV\ \eqref{eq:p5} is that the monodromy parameters $\th_{0}$, $\th_{1}$ and $\th_{\infty}$ are changed by integers.

Fokas, Mu\u{g}an and Ablowitz \cite{refFMA} show that the Schlesinger transformations for \PV\ \eqref{eq:p5} are in three sets
\[ (\mathrm{i}):\enskip \left\{ \begin{array}{lll} \th_{0}&\to&\th_{0}\pm1,\\ \th_{1} &\to&\th_{1},\\ \th_{\infty} &\to&\th_{\infty}\pm1, \end{array}\right. \qquad
(\mathrm{ii}):\enskip \left\{ \begin{array}{lll} \th_{0} &\to&\th_{0},\\ \th_{1} &\to&\th_{1}\pm1,\\ \th_{\infty} &\to&\th_{\infty}\pm1, \end{array}\right. \qquad
(\mathrm{iii}):\enskip \left\{ \begin{array}{lll} \th_{0} &\to&\th_{0}\pm1,\\ \th_{1} &\to&\th_{1}\pm1,\\ \th_{\infty} &\to&\th_{\infty}, \end{array}\right.\]
where the $\pm$ signs are independent.
Fokas, Mu\u{g}an and Ablowitz \cite[equations (4.44)]{refFMA} derive the eight Schlesinger transformations $\mathcal{R}_{1}$, $\mathcal{R}_{2}, \ldots,\mathcal{R}_{8}$, 
%$\mathcal{R}_{j}$, $j=1,2,\ldots,8$, 
in the first two sets where the monodromy parameter $\th_{\infty}$ changes by $\pm1$ and either $\th_{0}$ or $\th_{1}$ change by $\pm1$; see also \cite[equations (5.35)--(5.42)]{refMF}. 

Our interest is the effect of the Schlesinger transformations $\mathcal{R}_{1},\mathcal{R}_{2},\ldots,\mathcal{R}_{8}$ on the parameters $(\a,\b,\ga)$ and solutions $w(z)$ of \PV\ \eqref{eq:p5}. 
Using \eqref{SchlesParams} it is straightforward to compute what happens to $\a$, $\b$ and $\ga$ when $\th_{0}$, $\th_{1}$ and $\th_{\infty}$ change by $\pm1$. Hence using the results in Tables \ref{btab}, \ref{btac} and \ref{btbc} we can deduce the combination of \bts\ $\Tr_{\ep_{1},\ep_{2},\ep_{3}}$ \eqref{eq:PT.BTs.P5} that are equivalent to each of the Schlesinger transformations $\mathcal{R}_{1},\mathcal{R}_{2},\ldots,\mathcal{R}_{8}$. 
These results are given in Table \ref{Sch:R1R8}. Consequently it is straightforward to calculate the effect of the Schlesinger transformations $\mathcal{R}_{1},\mathcal{R}_{2},\ldots,\mathcal{R}_{8}$ to a solution $w(z)$ of \PV\ \eqref{eq:p5} with parameters $\pms{\hf a^2}{-\hf b^2}{c}$. These are given in equations \eqref{SchR1R8w} which are transformations that are quadratic in ${\rmd w}/{\rmd z}$.

\begin{table}[ht]
\[\begin{array}{|c|ccc|ccc|c|}
\hline
%& \th_{0}' & \th_{1}' & \th_{\infty}' & \a_{j} & \b_{j} & \ga_{j} & \\ \hline
& \th_{0} & \th_{1} & \th_{\infty} & \a & \b & \ga & \\ \hline
\mathcal{R}_{1} & \th_{0}+1 & \th_{1} & \th_{\infty}+1 & \hf(a+1)^2 & -\hf b^2 & c-1 & \Tr_{1,1,1}^{\,2}\circ\Tr_{-1,-1,-1}\\
\mathcal{R}_{2} & \th_{0}-1 & \th_{1} & \th_{\infty}-1 & \hf(a-1)^2 & -\hf b^2 & c+1& \Tr_{-1,-1,-1}^{\,2}\circ\Tr_{1,1,1} \\
\mathcal{R}_{3} & \th_{0}+1 & \th_{1} & \th_{\infty}-1 & \hf a^2 & -\hf(b+1)^2 & c-1& \Tr_{-1,1,1}^{\,2}\circ\Tr_{1,-1,-1}\\
\mathcal{R}_{4} & \th_{0}-1 & \th_{1} & \th_{\infty}+1 & \hf a^2 & -\hf(b-1)^2 & c+1& \Tr_{1,-1,-1}^{\,2}\circ\Tr_{-1,1,1}\\
\mathcal{R}_{5} & \th_{0} & \th_{1} +1& \th_{\infty}+1 & \hf a^2 & -\hf(b-1)^2 & c-1& \Tr_{1,1,1}^{\,3}\\
\mathcal{R}_{6} & \th_{0} & \th_{1}-1 & \th_{\infty}-1 & \hf a^2 & -\hf(b+1)^2 & c+1& \Tr_{-1,-1,-1}^{\,3}\\
\mathcal{R}_{7} & \th_{0} & \th_{1} +1& \th_{\infty} -1 & \hf(a-1)^2 & -\hf b^2 & c-1 &\Tr_{-1,1,1}^{\,3}\\
\mathcal{R}_{8} & \th_{0} & \th_{1}-1 & \th_{\infty} +1 & \hf(a+1)^2 & -\hf b^2 & c+1& \Tr_{1,-1,-1}^{\,3}\\ 
\hline 
\end{array}\]
\caption{\label{Sch:R1R8}The effect of the Schlesinger transformations %$\mathcal{R}_{j}$, $j=1,2,\ldots,8$ 
$\mathcal{R}_{1},\mathcal{R}_{2},\ldots,\mathcal{R}_{8}$ on the parameters $\th_{0}$, $\th_{1}$, $\th_{\infty}$, $\a$, $\b$ and $\ga$ and the equivalent combination of \bts\ $\Tr_{\ep_{1},\ep_{2},\ep_{3}}$ \eqref{eq:PT.BTs.P5}.}
\end{table}

\begin{subequations}\label{SchR1R8w}\begin{align}
\mathcal{R}_{1}:\quad w_{1}&=\dfrac{\big\{zw'+aw^2-(a+b+z)w+b\big\}\big\{zw'+aw^2-(a-b+z)w-b\big\}}{\big\{zw'+ a w(w -1)\big\}^{\!2}-z^2w^2-2z(a-c+1)w(w-1)-b^2(w-1)^2},\\
%\mathcal{R}_{1}:\quad w_{1}&=\dfrac{\big\{zw'+aw^2-(a+b+z)w+b\big\}\big\{zw'+aw^2-(a-b+z)w-b\big\}}{z^{2}\big(w'\big)^{\!2} -2 a z w(w -1)w'-z^2w^2-2z(a-c+1)w(w-1)+(a^2w^2-b^2)(w-1)^2},\\
%\mathcal{R}_{2}:\quad w_{2}&=\dfrac{\big\{zw'-aw^2+(a+b+z)w-b\big\}\big\{zw'-aw^2+(a-b+z)w+b\big\}}{z^{2}\big(w'\big)^{\!2}+2 a z w(w -1)w'-z^2w^2-2z(a-c-1)w(w-1)+(a^2w^2-b^2)(w-1)^2},\\
\mathcal{R}_{2}:\quad w_{2}&=\dfrac{\big\{zw'-aw^2+(a+b+z)w-b\big\}\big\{zw'-aw^2+(a-b+z)w+b\big\}}{\big\{zw'- aw(w -1)\big\}^{\!2}-z^2w^2-2z(a-c-1)w(w-1)-b^2(w-1)^2},\\
\mathcal{R}_{3}:\quad w_{3}&=1+\frac{1}{a}\left\{\frac{(a+b-c+1)zw}{zw'+aw^2-(a-b+z)w-b} +\frac{(a-b+c-1)zw}{zw'-aw^2+(a+b-z)w-b}\right\},\\
\mathcal{R}_{4}:\quad w_{4}&=1-\frac{1}{a}\left\{\frac{(a-b+c+1)zw}{zw'+aw^2-(a+b-z)w+b} +\frac{(a+b-c-1)zw}{zw'-aw^2+(a-b+z)w+b}\right\},\\
\mathcal{R}_{5}:\quad w_{5}&=1+\frac{1}{a}\left\{\frac{(a-b-c+1)zw}{zw'+aw^2-(a+b+z)w+b} +\frac{(a+b+c-1)zw}{zw'-aw^2+(a-b-z)w+b}\right\},\\
\mathcal{R}_{6}:\quad w_{6}&=1-\frac{1}{a}\left\{\frac{(a+b+c+1)zw}{zw'+aw^2-(a-b-z)w-b} +\frac{(a-b-c-1)zw}{zw'-aw^2+(a+b+z)w-b}\right\},\\
\mathcal{R}_{7}:\quad w_{7}&=\frac{\big\{zw'-aw^2+(a-b-z)w+b \big\} \big\{zw'-aw^2+(a+b-z)w-b \big\}}{\big\{zw'-aw(w-1)\big\}^{\!2}-z^2w^2+2(a+c-1)zw(w-1)-b^2(w-1)^2},\\
%\mathcal{R}_{7}:\quad w_{7}&=\frac{\big\{zw'-aw^2+(a-b-z)w+b \big\} \big\{zw'-aw^2+(a+b-z)w-b \big\}}{z^{2}\big(w'\big)^{\!2}-2azw(w-1)w'-z^2w^2+2(a+c-1)zw(w-1)+(a^2w^2-b^2)(w-1)^2},\\
%\mathcal{R}_{8}:\quad w_{8}&=\frac{\big\{zw'+aw^2-(a-b-z)w-b \big\} \big\{zw'+aw^2-(a+b-z)w+b \big\}}{z^{2}\big(w'\big)^{\!2}+2azw(w-1)w'-z^2w^2+2(a+c+1)zw(w-1)+(a^2w^2-b^2)(w-1)^2},
\mathcal{R}_{8}:\quad w_{8}&=\frac{\big\{zw'+aw^2-(a-b-z)w-b \big\} \big\{zw'+aw^2-(a+b-z)w+b \big\}}{\big\{zw'+aw(w-1)\big\}^{\!2}-z^2w^2+2(a+c+1)zw(w-1)-b^2(w-1)^2},
\end{align}\end{subequations}
with $'=\rmd/\rmd z$.

\begin{lemma}\label{lemma28}The Schlesinger transformations $\mathcal{R}_{1},\mathcal{R}_{2},\ldots,\mathcal{R}_{8}$ commute pairwise, i.e.
\[ \mathcal{R}_{j}\mathcal{R}_{k}=\mathcal{R}_{k}\mathcal{R}_{j},\qquad j,k,=1,2,\ldots,8.\]
\end{lemma}
\begin{proof}It is straightforward to verify this using the definition of the transformations given in Table \ref{Sch:R1R8} and \eqref{SchR1R8w}.
\end{proof}

The Schlesinger transformations $\mathcal{R}_{9},\ldots,\mathcal{R}_{12}$ in which $\th_{0}$ and $\th_{1}$ change by $\pm1$, with $\th_{\infty}$ unchanged, are obtained using a combination of two Schlesinger transformations as follows
\beq\label{def:R9R12}
\mathcal{R}_{9}=\mathcal{R}_{1}\mathcal{R}_{7},\qquad\mathcal{R}_{10}=\mathcal{R}_{2}\mathcal{R}_{8},\qquad\mathcal{R}_{11}=\mathcal{R}_{1}\mathcal{R}_{6},\qquad\mathcal{R}_{12}=\mathcal{R}_{2}\mathcal{R}_{5}.\eeq
Due to Lemma \ref{lemma28}, the order of the Schlesinger transformations in \eqref{def:R9R12} is irrelevant.
The effect of the Schlesinger transformations $\mathcal{R}_{9},\ldots,\mathcal{R}_{12}$ on the parameters $\th_{0}$, $\th_{1}$, $\th_{\infty}$, $\a$, $\b$ and $\ga$ and the equivalent combination of \bts\ $\Tr_{\ep_{1},\ep_{2},\ep_{3}}$ \eqref{eq:PT.BTs.P5} is given in Table \ref{Sch:R9R14}. However, from Table \ref{Sch:R9R14} we see that for the parameters $a$ and $b$ in $\a$ and $\b$ respectively to change in opposite directions, then it is necessary to introduce the transformations $\mathcal{R}_{13}$ and $\mathcal{R}_{14}$, which can be done in two ways
\beq\label{def:R13R14 } \mathcal{R}_{13}=\mathcal{R}_{1}\mathcal{R}_{4}=\mathcal{R}_{5}\mathcal{R}_{8},\qquad \mathcal{R}_{14}=\mathcal{R}_{2}\mathcal{R}_{3}=\mathcal{R}_{6}\mathcal{R}_{7}, \eeq
and their effect is also given in Table \ref{Sch:R9R14}. The effect of the Schlesinger transformations $\mathcal{R}_{9},\mathcal{R}_{10},\ldots,\mathcal{R}_{14}$ to a solution $w(z)$ of \PV\ \eqref{eq:p5} with parameters $\pms{\hf a^2}{-\hf b^2}{c}$ is given in equations \eqref{SchR9R14w} which are also transformations that are quadratic in ${\rmd w}/{\rmd z}$.

\begin{table}[ht]
\[\begin{array}{|c|ccc|ccc|c|}
\hline
%& \th_{0}' & \th_{1}' & \th_{\infty}' & \a_{j} & \b_{j} & \ga_{j} & \\ \hline
& \th_{0} & \th_{1} & \th_{\infty} & \a & \b & \ga & \\ \hline
\mathcal{R}_{9}=\mathcal{R}_{1}\mathcal{R}_{7} & \th_{0}+1 & \th_{1}+1 & \th_{\infty} & \hf a^2 & -\hf b^2 & c-2& \Tr_{-1,1,1}^{\,3}\circ\Tr_{1,1,1}^{\,2}\circ\Tr_{-1,-1,-1}\\ 
\mathcal{R}_{10}=\mathcal{R}_{2}\mathcal{R}_{8} & \th_{0}-1 & \th_{1}-1 & \th_{\infty} & \hf a^2 & -\hf b^2 & c+2& \Tr_{1,-1,-1}^{\,3}\circ \Tr_{-1,-1,-1}^{\,2}\circ\Tr_{1,1,1} \\ 
\mathcal{R}_{11}=\mathcal{R}_{1}\mathcal{R}_{6} & \th_{0}+1 & \th_{1}-1 & \th_{\infty} & \hf(a+1)^2 & -\hf(b+1)^2 & c& \Tr_{-1,1,-1}\circ\Tr_{-1,-1,-1} \\
%\Tr_{1,-1,1}\circ \Tr_{-1,1,1}\circ \Tr_{1,-1,1}\circ \Tr_{1,-1,-1}\\ 
\mathcal{R}_{12}=\mathcal{R}_{2}\mathcal{R}_{5} & \th_{0}-1 & \th_{1}+1 & \th_{\infty} & \hf(a-1)^2 & -\hf(b-1)^2 & c& \Tr_{1,-1,1}\circ\Tr_{1,1,1} \\ \hline
% \Tr_{-1,1,-1}\circ \Tr_{1,-1,-1}\circ \Tr_{-1,1,-1}\circ \Tr_{-1,1,1}\\ 
\mathcal{R}_{13}=\mathcal{R}_{1}\mathcal{R}_{4}& \th_{0} & \th_{1} & \th_{\infty}+2 & \hf(a+1)^2 & -\hf(b-1)^2 & c&\Tr_{1,-1,1}^{\,2}\\ 
%\mathcal{R}_{5}\mathcal{R}_{8}& \th_{0} & \th_{1} & \th_{\infty}+2 & \hf(a+1)^2 & -\hf(b-1)^2 & c& \Tr_{1,-1,1}^{\,2}\\ 
\mathcal{R}_{14}=\mathcal{R}_{2}\mathcal{R}_{3}& \th_{0} & \th_{1} & \th_{\infty}-2 & \hf(a-1)^2 & -\hf(b+1)^2 & c& \Tr_{-1,1,-1}^{\,2}\\ 
%\mathcal{R}_{6}\mathcal{R}_{7}& \th_{0} & \th_{1} & \th_{\infty}-2 & \hf(a-1)^2 & -\hf(b+1)^2 & c& \Tr_{-1,1,-1}^{\,2}\\ 
\hline
\end{array}\]
\caption{\label{Sch:R9R14}The effect of the Schlesinger transformations $\mathcal{R}_{9},\mathcal{R}_{10},\ldots,\mathcal{R}_{14}$ on the parameters $\th_{0}$, $\th_{1}$, $\th_{\infty}$, $\a$, $\b$ and $\ga$ and the equivalent combination of \bts\ $\Tr_{\ep_{1},\ep_{2},\ep_{3}}$ \eqref{eq:PT.BTs.P5}.}
\end{table}

\begin{subequations}\label{SchR9R14w}\begin{align}
%w_{9}&=\frac{\big[zw'-(b-c+1)w^2+(2b-c-z+1)w-b \big]\big[ zw'+(b+c-1)w^2-(2b+c+z-1)w+b\big]}{w\big[zw'-aw^2+(2a+c-z-1)w-a-c+1\big]\big[zw'+aw^2-(2a-c+z+1)w+a-c+1 \big]},\\
\mathcal{R}_{9}:\enskip w_{9}&=\frac{1}{w}+\frac{(w-1)^2}{2aw}\left\{\frac{(a+b+c-1)(a-b+c-1)}{zw'-aw^2+(2a+c-z-1)w-a-c+1} \right. \nonumber\\ &\qquad\qquad\qquad\qquad\qquad \left. 
-\frac{(a+b-c+1)(a-b-c+1)}{zw'+aw^2-(2a-c+z+1)w+a-c+1} \right\},\\
\mathcal{R}_{10}:\enskip w_{10}&=\frac{1}{w}+\frac{(w-1)^2}{2aw}\left\{\frac{(a+b-c-1)(a-b-c-1)}{zw'-aw^2+(2a-c+z-1)w-a+c+1} \right. \nonumber \\ &\qquad\qquad\qquad\qquad\qquad -\left. 
\frac{(a+b+c+1)(a-b+c+1)}{zw'+aw^2-(2a+c-z+1)w+a+c+1} \right\},\\
%w_{10}&=\frac{\big[zw'+(b-c-1)w^2-(2b-c-z-1)w+b \big]\big[ zw'-(b+c+1)w^2+(2b+c+z+1)w-b\big]}{w\big[zw'-aw^2+(2a-c+z-1)w-a+c+1\big]\big[zw'+aw^2-(2a+c-z+1)w+a+c+1 \big]},\\
\mathcal{R}_{11}:\enskip w_{11}&=\frac{\big[zw'-(w-1)(bw-b+w)\big]^{\!2}-z^2w^2+2cz(w-1)w^2-(a+b+1)^2(w-1)^2w^2}{w\Big\{\big[zw'+(w-1)(aw-a-1)\big]^{\!2}-z^2w^2+2cz(w-1)w-(a+b+1)^2(w-1)^2\Big\}},\\
\mathcal{R}_{12}:\enskip w_{12}&=\frac{\big[zw'+(w-1)(bw-b-w)\big]^{\!2}-z^2w^2+2cz(w-1)w^2-(a+b-1)^2(w-1)^2w^2}{w\Big\{\big[zw'-(w-1)(aw-a+1)\big]^{\!2}-z^2w^2+2cz(w-1)w-(a+b-1)^2(w-1)^2\Big\}},\\
\mathcal{R}_{13}:\enskip w_{13}&=\frac{\big[zw'+(w-1)(bw-b-w)\big]^{\!2}-z^2w^2+2cz(w-1)w^2-(a-b+1)^2(w-1)^2w^2}{w\Big\{\big[zw'+(w-1)(aw-a-1)\big]^{\!2}-z^2w^2+2cz(w-1)w-(a-b+1)^2(w-1)^2\Big\}},\\
\mathcal{R}_{14}:\enskip w_{14}&=\frac{\big[zw'-(w-1)(bw-b+w)\big]^{\!2}-z^2w^2+2cz(w-1)w^2-(a-b-1)^2(w-1)^2w^2}{w\Big\{\big[zw'-(w-1)(aw-a+1)\big]^{\!2}-z^2w^2+2cz(w-1)w-(a-b-1)^2(w-1)^2\Big\}}.
\end{align}\end{subequations}

\subsection{\label{ssec:spv}\bts\ of the symmetric \PV}
From the works of Okamoto
\cite{refOkamotoP2P4,refOkamotoPV,refOkamotoPVI,refOkamotoPIII}, it is known that the parameter spaces of \PII--\PVI\ all admit the action of an
extended affine Weyl group; the group acts as a group of \bts. 
In a series of papers, Noumi and Yamada \cite{refNoumi,refNY98i,refNY98iii,refNY04}
have implemented this idea to derive a hierarchy of dynamical systems
associated to the affine Weyl group of type $\A{N}$, often known as ``\textit{symmetric forms of the \peqs}". The behaviour of each dynamical
system varies depending on whether $N$ is even or odd. 
The first member of the $\A{2n+1}$ hierarchy, i.e.\ $\A{3}$, usually known as symmetric-\PV\ (\spv), is equivalent to \PV\ \eqref{eq:p5} and given by
\begin{subequations}\label{eq:spv}\begin{align}
z\deriv{f_{0}}{z}&=f_{0}f_{2}(f_{1}-f_{3})+(\hf-\nu_{2})f_{0}+\nu_{0}f_{2},\\
z\deriv{f_{1}}{z}&=f_{1}f_{3}(f_{2}-f_{0})+(\hf-\nu_{3})f_{1}+\nu_{1}f_{3},\\
z\deriv{f_{2}}{z}&=f_{2}f_{0}(f_{3}-f_{1})+(\hf-\nu_{0})f_{2}+\nu_{2}f_{0},\\
z\deriv{f_{3}}{z}&=f_{3}f_{1}(f_{0}-f_{2})+(\hf-\nu_{1})f_{3}+\nu_{3}f_{1},
\end{align}
with the normalisations
\beq\label{eq:sPVa}
%\nu_{0}+\nu_{1}+\nu_{2}+\nu_{3}=1,\qquad 
f_{0}(z)+f_{2}(z)=\sqrt{z},\qquad f_{1}(z)+f_{3}(z)=\sqrt{z},
\eeq
and $\nu_{0}$, $\nu_{1}$, $\nu_{2}$ and $\nu_{3}$ are constants such that
\beq\label{eq:sPVk}
\nu_{0}+\nu_{1}+\nu_{2}+\nu_{3}=1.
\eeq\end{subequations}
Here $w=-f_{2}/f_{0}$ satisfies \PV\ \eqref{eq:p5} with $\pms{\hf\nu_{0}^2}{-\hf\nu_{2}^2}{\nu_{3}-\nu_{1}}$ and so
\begin{align*}
f_0 &= \frac{\sqrt z}{1-w}, \qquad
f_1 = \frac{1}{2\sqrt z\,w} \left\{z w - z\deriv{w}{z}+ (w-1)(\nu_0 w+\nu_2)\right\}, \\
f_2 &= \frac{\sqrt z\,w}{w-1}, \qquad
f_3 = \frac{1}{2\sqrt z\,w} \left\{z w + z \deriv{w}{z} - (w-1)(\nu_0 w+\nu_2)\right\}.
\end{align*}
%, or alternatively $w=-f_{3}/f_{1}$ satisfies \PV\ with $\pms{\hf\nu_{1}^2}{-\hf\nu_{3}^2}{\nu_{0}-\nu_{2}}$.
The \spv\ \eqref{eq:spv} was found by Adler \cite{refAdler} in the context of periodic chains of \bts\ and has applications in random matrix theory, cf.~\cite{refFW02}. 

The \bts\ of \spv\ \eqref{eq:spv} are given by the extended affine Weyl group of type $A_{3}^{(1)}$ \cite{refAdler,refNY98iii} which has generators $\langle\bold{s}_{0},\bold{s}_{1},\bold{s}_{2},\bold{s}_{3},\bold{\pi}\rangle$ that satisfy the algebraic relations
\begin{align*}&\bold{s}_{j}^2=\mathcal{I},\qquad \bold{s}_{j}\bold{s}_{j+2}=\bold{s}_{j+2}\bold{s}_{j},\qquad \bold{s}_{j}\bold{s}_{j+1}\bold{s}_{j}=\bold{s}_{j+1}\bold{s}_{j}\bold{s}_{j+1},\\
&\bold{\pi}^4=\mathcal{I},\qquad \bold{\pi}\bold{s}_{j} =\bold{s}_{j+1}\bold{\pi}.\end{align*}
The reflections $\bold{s}_{j}$ act on the parameters by
\[\bold{s}_{j}(\nu_{j})=-\nu_{j},\qquad\bold{s}_{j}(\nu_{j\pm1})=\nu_{j\pm1}+\nu_{j},\qquad
\bold{s}_{j}(\nu_{j+2})=\nu_{j+2},\] and on the variables by
\[\bold{s}_{j}(f_{j})=f_{j},\qquad\bold{s}_{j}(f_{j\pm1})=f_{j\pm1}\pm\frac{\nu_{j}}{f_{j}},\qquad\bold{s}_{j}(f_{j+2})=f_{j+2}. \]
The rotation $\bold{\pi}$ acts
\[ \bold{\pi}(\nu_{j})=\nu_{j+1},\qquad \bold{\pi}(f_{j})=f_{j+1}.\]
These \bts\ of \spv\ \eqref{eq:spv} are summarised in Table \ref{table26}.

\comment{\[\begin{array}{|c|cccc|cc|} \hline
& \nu_{0} & \nu_{1} & \nu_{2} & \nu_{3} & q & p \\ \hline
\bold{s}_{0} & -\nu_{0} & \nu_{1}+\nu_{0} & \nu_{2} & \nu_{3}+\nu_{0} & q+\dfrac{\nu_{0}}{p+z} & p \\
\bold{s}_{1} & \nu_{0}+\nu_{1} & -\nu_{1} & \nu_{2}+\nu_{1} & \nu_{3} & q & p-\dfrac{\nu_{1}}{q}\\
\bold{s}_{2} & \nu_{0} & \nu_{1}+\nu_{2} & -\nu_{2} & \nu_{3}+\nu_{2} & q +\dfrac{\nu_{2}}{p} & p \\
\bold{s}_{3} & \nu_{0}+\nu_{3} & \nu_{1} & \nu_{2}+\nu_{3} & -\nu_{3} & q & p-\dfrac{\nu_{3}}{q-1} \\ 
\bold{\pi} & \nu_{1} & \nu_{2} & \nu_{3} & \nu_{0} & -\dfrac{p}{z} & z(q-1) \\ \hline
\end{array} \]}
\comment{\begin{subequations}\begin{align} 
\bold{s}_{i}(\nu_{j})=\nu_{j}-a_{ij}\nu_{i},\qquad \pi(\nu_{j})=\nu_{j+1},\\
\bold{s}_{i}(f_{j})=f_{j}+u_{ij}\frac{\nu_{i}}{f_{i}},\qquad \pi(f_{j})=f_{j+1},
\end{align}\end{subequations}
\[ \Matrix{A}=\left(\begin{matrix} 2 & -1 & 0 & -1\\ -1 & 2 & -1 & 0 \\ 0 & -1 & 2 & -1 \\ -1 & 0 & -1 & 2\end{matrix}\right),\qquad
\Matrix{U}=\left(\begin{matrix} 0 & 1 & 0 & -1\\ -1 & 0 & 1 & 0 \\ 0 & -1 & 0 & 1 \\ 1 & 0 & -1 & 0\end{matrix}\right),\]}

\begin{table}[ht]
\[\begin{array}{|c|cccc|cccc|} \hline
& \nu_{0} & \nu_{1} & \nu_{2} & \nu_{3} & f_{0} & f_{1} & f_{2} & f_{3} \\ \hline
\bold{s}_{0} & -\nu_{0} & \nu_{1}+\nu_{0} & \nu_{2} & \nu_{3}+\nu_{0} & f_{0} & f_{1}+\dfrac{\nu_{0}}{f_{0}} & f_{2} & f_{3} -\dfrac{\nu_{0}}{f_{0}} \\
\bold{s}_{1} & \nu_{0}+\nu_{1} & -\nu_{1} & \nu_{2}+\nu_{1} & \nu_{3} & f_{0}-\dfrac{\nu_{1}}{f_{1}} & f_{1} & f_{2} +\dfrac{\nu_{1}}{f_{1}} & f_{3}\\
\bold{s}_{2} & \nu_{0} & \nu_{1}+\nu_{2} & -\nu_{2} & \nu_{3}+\nu_{2} & f_{0}& f_{1} -\dfrac{\nu_{2}}{f_{2}} & f_{2} & f_{3} +\dfrac{\nu_{2}}{f_{2}} \\
\bold{s}_{3} & \nu_{0}+\nu_{3} & \nu_{1} & \nu_{2}+\nu_{3} & -\nu_{3} & f_{0}+\dfrac{\nu_{3}}{f_{3}} & f_{1} & f_{2}-\dfrac{\nu_{3}}{f_{3}} & f_{3} \\ 
\bold{\pi} & \nu_{1} & \nu_{2} & \nu_{3} & \nu_{0} & f_{1} & f_{2} & f_{3} & f_{0} \\ \hline
\end{array} \]
\caption{\label{table26}The \bts\ of \spv\ \eqref{eq:spv}.}
\end{table}

Masuda, Ohta and Kajiwara \cite{refMOK} define the translation operators $\Top_{j}$, $j=0,1,2,3$, given by
\beq \Top_{0}= \bold{s}_{3} \bold{s}_{2} \bold{s}_{1} \bold{\pi},\qquad \Top_{1}= \bold{\pi} \bold{s}_{3} \bold{s}_{2} \bold{s}_{1},\qquad \Top_{2}= \bold{s}_{1} \bold{\pi} \bold{s}_{3} \bold{s}_{2},\qquad \Top_{3}= \bold{s}_{2} \bold{s}_{1} \bold{\pi} \bold{s}_{3},\label{Toperator}\eeq
which commute with each other and satisfy $\Top_{0}\Top_{1}\Top_{2}\Top_{3}=\mathcal{I}$, the identity. The effect of the translation operators $\Top_{j}$ %, $j=0,1,2,3$ 
on the parameters $\nu_{0}$, $\nu_{1}$, $\nu_{2}$, $\nu_{3}$, $a$, $b$ and $c$ is given in Table \ref{table27}. Since the parameters change by $\pm1$ then using the results in Tables \ref{btab}, \ref{btac} and \ref{btbc} we can deduce the combination of \bts\ $\Tr_{\ep_{1},\ep_{2},\ep_{3}}$ \eqref{eq:PT.BTs.P5} that are equivalent to the translation operators $\Top_{j}$ \eqref{Toperator}. %, $j=0,1,2,3$  
The equivalent Schlesinger transformations $\mathcal{R}_{j}$ using Table \ref{Sch:R1R8} are also given in Table \ref{table27}. 
\comment{By comparing Tables \ref{Sch:R1R8} and \ref{table27}, we see that the translation operators given by \eqref{Toperator} are equivalent to Schlesinger transformations 
$\mathcal{R}_{1},\mathcal{R}_{2},\ldots,\mathcal{R}_{8}$, as follows
\[\begin{array}{l@{\qquad}l@{\qquad}l@{\qquad}l}
\Top_{0}=\mathcal{R}_{3},& \Top_{1}=\mathcal{R}_{2},& \Top_{2}=\mathcal{R}_{8},& \Top_{3}=\mathcal{R}_{5}, \\
\Top_{0}^{-1}=\mathcal{R}_{4},& \Top_{1}^{-1}=\mathcal{R}_{1},& \Top_{2}^{-1}=\mathcal{R}_{7},& \Top_{3}^{-1}=\mathcal{R}_{6}.
\end{array}\]}

\begin{table}[ht]
\[\begin{array}{|c|cccc|ccc|c|c|} \hline
& \nu_{0} & \nu_{1} & \nu_{2} & \nu_{3} & a & b & c & & \\ \hline
\Top_{0} & \nu_{0}-1 & \nu_{1} & \nu_{2} & \nu_{3}+1 & a & b+1 & c-1 &\Tr_{-1,1,1}^{\,2}\circ\Tr_{1,-1,-1} & \mathcal{R}_{3}\\ 
\Top_{1} & \nu_{0} +1& \nu_{1} -1& \nu_{2} & \nu_{3} & a-1 & b & c+1 &\Tr_{-1,-1,-1}^{\,2}\circ\Tr_{1,1,1} & \mathcal{R}_{2}\\ 
\Top_{2} & \nu_{0} & \nu_{1}+1 & \nu_{2} -1& \nu_{3} & a+1 & b & c +1 & \Tr_{1,-1,-1}^{\,3} & \mathcal{R}_{8}\\ 
\Top_{3} & \nu_{0} & \nu_{1} & \nu_{2}+1& \nu_{3} -1 & a & b-1 & c-1 &\Tr_{1,1,1}^{\,3} & \mathcal{R}_{5}\\ \hline
\Top_{0}^{-1} & \nu_{0}+1 & \nu_{1} & \nu_{2} & \nu_{3}-1 & a & b-1 & c+1 &\Tr_{1,-1,-1}^{\,2}\circ\Tr_{-1,1,1} & \mathcal{R}_{4}\\ 
\Top_{1}^{-1} & \nu_{0} -1& \nu_{1} +1& \nu_{2} & \nu_{3} & a+1 & b & c-1 &\Tr_{1,1,1}^{\,2}\circ\Tr_{-1,-1,-1} & \mathcal{R}_{1}\\ 
\Top_{2}^{-1} & \nu_{0} & \nu_{1}-1 & \nu_{2} +1& \nu_{3} & a-1 & b & c -1 & \Tr_{-1,1,1}^{3} & \mathcal{R}_{7} \\ 
\Top_{3}^{-1} & \nu_{0} & \nu_{1} & \nu_{2}-1& \nu_{3} +1 & a & b+1 & c+1 &\Tr_{-1,-1,-1}^{\,3} & \mathcal{R}_{6} \\ \hline
\end{array} \]
\caption{\label{table27}The effect of the translation operators $\Top_{j}$ 
%$j=0,1,2,3$ given by 
\eqref{Toperator} on the parameters $\nu_{0}$, $\nu_{1}$, $\nu_{2}$, $\nu_{3}$, $a$, $b$ and $c$ together with the equivalent combination of \bts\ $\Tr_{\ep_{1},\ep_{2},\ep_{3}}$ \eqref{eq:PT.BTs.P5} and Schlesinger transformation $\mathcal{R}_{j}$.}
\end{table}

From the Hamiltonian \cite{refKMNOY,refWat98}
\beq \label{p5:ham2} \mathcal{H}_\mathrm{V}(q,p,z;\nu_{0},\nu_{1},\nu_{2},\nu_{3})=q(q-1)p(p+z)-\nu_{3}(q-1)p+\nu_{0}zq-\nu_{1}qp,\eeq
with $\nu_{0}+\nu_{1}+\nu_{2}+\nu_{3}=1$,
we obtain the system
\begin{subequations}\label{sys:qp}\begin{align} 
z\deriv{q}{z}&=\pderiv{\mathcal{H}_\mathrm{V}}{p}=q(q-1)(2p+z)-\nu_{3}(q-1)-\nu_{1}q,\label{sys:qp1}\\
z\deriv{p}{z}&=-\pderiv{\mathcal{H}_\mathrm{V}}{q}=-(2q-1)p(p+z)+(\nu_{1}+\nu_{3})p-\nu_{0}z.\label{sys:qp2}
\end{align}\end{subequations}
Eliminating $p$ in \eqref{sys:qp} gives
\beq \deriv[2]{q}{z}= %\frac{1}{2}\left(\frac{1}{q}+\frac{1}{q-1}\right) %
\frac{2q-1}{2q(q-1)}\!\left(\deriv{q}{z}\right)^{\!2} -\frac{1}{z}\deriv{q}{z}+\frac{\nu_{3}^2(q-1)^2-\nu_{1}^2q^2}{2z^2q(q-1)}
+\frac{(\nu_{2}-\nu_{0})q(q-1)}{z}+\hf q(q-1)(2q-1),\eeq
and then letting $q=w/(w-1)$ yields \PV\ \eqref{eq:p5} with $\pms{\hf\nu_{1}^2}{-\hf\nu_{3}^2}{\nu_{0}-\nu_{2}}$.
%and then letting $q=1/(1-w)$ yields \PV\ \eqref{eq:p5} with $\pms{\hf\nu_{3}^2}{-\hf\nu_{1}^2}{\nu_{2}-\nu_{0}}$.
Similarly, eliminating $q$ in \eqref{sys:qp} gives
\begin{align} \deriv[2]{p}{z}&=\frac{2p+z}{2p(p+z)}\!\left(\deriv{p}{z}\right)^{\!2}-\frac{p}{z(p+z)}\deriv{p}{z}+\frac{1-\nu_{2}^2}{2(p+z)} -\frac{\nu_{0}^2}{2p}\nonumber\\&\qquad
+\frac{(2p+3z+2\nu_{1}-2\nu_{3})p^2}{2z^2} +\frac{(z+2\nu_{1}-2\nu_{3})p-\nu_{0}^2+\nu_{2}^2-1}{2z},
\end{align}
and then letting $p=z/(w-1)$ yields \PV\ \eqref{eq:p5} with $\pms{\hf\nu_{0}^2}{-\hf\nu_{2}^2}{\nu_{3}-\nu_{1}}$.

\begin{remarks}\textrm{
\begin{enumerate} %\rmd w/\rmd z
\item The system \eqref{sys:qp} can be obtained from the \spv\ \eqref{eq:spv} through
\[f_{0}=-\frac{p}{\sqrt{z}},\qquad f_{1}=\sqrt{z}\,(1-q),\quad f_{2}=\frac{p+z}{\sqrt{z}},\qquad f_{3}=\sqrt{z}\,q.\]
\item We remark that \eqref{p5:ham2} is not the usual Hamiltonian for \PV\ \eqref{eq:p5}. Okamoto \cite{refOkamoto80a,refOkamoto80b,refOkamotoPV} derived the Hamiltonian 
\beq\label{HJM:47}
\mathcal{H}_\mathrm{V}^{\rm Ok}(Q,P,z) = Q(Q-1)^2P^2 -\left\{b(Q-1)^2-(c+1)Q(Q-1)+zQ\right\}P+\k(Q-1),\eeq
where $\k=\tfrac{1}{4}(b-c-1)^2-\tfrac{1}{4}a^2$,
with $a$, $b$ and $c$ parameters. Consequently we obtain the system
\begin{subequations}\label{HJM:4950}
\begin{align}\label{HJM:49}
z\deriv{Q}{z}&=\pderiv{\mathcal{H}_\mathrm{V}^{\rm Ok}}{P}=2Q(Q-1)^2P-b(Q-1)^2+(c+1)Q(Q-1)-zQ,\\
z\deriv{P}{z}&=-\pderiv{\mathcal{H}_\mathrm{V}^{\rm Ok}}{Q}=-(3Q-1)(Q-1)P^2+2(b-c-1)QP+(z-2b+c+1)P%+\tfrac{1}{4}\big\{a^2-(b-c-1)^2\big\}
-\k. \label{HJM:50}
\end{align}\end{subequations}
Eliminating $P$ then $Q=w$ satisfies \PV\ \eqref{eq:p5}
with $\pms{\hf a^2}{-\hf b^2}{c}$. The Hamiltonians $\mathcal{H}_\mathrm{V}$ \eqref{p5:ham2} and $\mathcal{H}_\mathrm{V}^{\rm Ok}$ \eqref{HJM:47} are related
\[ \mathcal{H}_\mathrm{V}(q,p,z)= \mathcal{H}_\mathrm{V}^{\rm Ok}(Q,P,z)+\tfrac{1}{4}(c+1)^2-\tfrac{1}{4}(a+b)^2,\]
by the transformation \cite{refWat95,refWat98}
\[ (q-1)(Q-1)=1,\qquad (q-1)p+(Q-1)P=\hf(a+b-c-1),\]
where the relationship between the parameters $\{\nu_{0},\nu_{1},\nu_{2},\nu_{3}\}$ and $\{a,b,c\}$ is given by
\[ \nu_{0}=-\hf(a+b-c-1),\qquad \nu_{1}=a,\qquad\nu_{2}=-\hf(a+b+c-1),\qquad\nu_{3}=b.\]
\end{enumerate}}\end{remarks}

\section{\label{sec:rats}Rational solutions}
In this section we discuss applications of the \bts\ described in the previous section to rational solutions of \PV\ \eqref{eq:p5}, which are classified in the following theorem.
\begin{theorem}\textrm{\label{thm:KLM}
\textit{Equation \eqref{eq:p5} has a rational solution if and only if one of the following holds:}
\begin{enumerate}[(i)]
\item\textit{$\a=\hf m^2$, $\b=-\hf(m+2n+1+\mu)^2$, $\ga=\mu$, for $m\geq1$;}
\item\textit{$\a=\hf(m+\mu)^2$, $\b=-\hf(n+\ep\mu)^2$, $\ga=m+\ep n$, with $\ep=\pm1$, provided that $m\not=0$ or $n\not=0$;}
\item\textit{$\a=\hf(m+\hf)^2$, $\b=-\hf(n+\hf)^2$, $\ga\notin\Integer$, provided that $m\not=0$ or $n\not=0$,}
\end{enumerate}
\textit{where $m,n\in\Integer$ and $\mu$ is an arbitrary constant,
together with the solutions obtained through the symmetries \eqref{PV:Sym}.}%$\mathcal{S}$ 
}\end{theorem}

\begin{proof}See Kitaev, Law and McLeod \cite[Theorem 1.1]{refKLM}; also \cite[Theorem 40.3]{refGLS}.\end{proof}

\begin{remarks}\textrm{
\begin{enumerate}
\item %Kitaev, Law and McLeod \cite[Theorem 1.1]{refKLM} give four cases, though their cases (I) and (II) are related by the symmetry \eqref{PV:Sym2}. Further 
Kitaev, Law and McLeod \cite[Theorem 1.2]{refKLM} prove that if $\ga\notin\Integer$ in case (i), then the rational solution is unique and if $\ga\in\Integer$ then there are at most two rational solutions. 
Examples of non-unique rational solutions of \PV\ \eqref{eq:p5} are given in \cite{refCD24}. %; see also \cite{refAGLZ24a,refAGLZ24b}.
In fact, the non-unique rational solutions are special cases of a special function solution of \PV\ \eqref{eq:p5} in terms of the functions $I_{n+1/2}(z)$, where $I_{\nu}(z)$ is the \textit{modified Bessel function} and $n\in\Integer$,
as we illustrated in \cite{refCDM26}. The functions $I_{n+1/2}(z)$, with $n\in\Integer$, are proportional to \textit{modified spherical Bessel functions}, cf.~\cite[\S10.47]{refDLMF}.
%\[{\mathsf{i}^{(1)}_{n}}\left(z\right)=\sqrt{\frac{\pi}{2z}}\,I_{n+\frac{1}{2}}(z)\] 
\item The requirement in case (iii) that $\ga\notin\Integer$ is not necessary as there exist rational solutions with $\ga\in\Integer$, see for example \eqref{sol:u3}. Further there are rational solutions in this case when $m=n=0$, see \eqref{sol:UmLag0}.
\item Rational solutions in case (i) are expressed in terms of \textit{generalised Laguerre polynomials}, which are discussed in \S\ref{ssec:GenLag}, and in cases (ii) and (iii) in terms of \textit{generalised Umemura polynomials}, discussed in \S\ref{ssec:GenUm}.
\end{enumerate}}\end{remarks}

\subsection{\label{ssec:GenLag}Generalised Laguerre polynomials}
Rational solutions in case (i) of Theorem\ \ref{thm:KLM} are expressed in terms of \textit{generalised Laguerre polynomials}, which are written in terms of a determinant of Laguerre polynomials, or equivalently as a Wronskian, properties of which are discussed in \cite{refBB,refCD24}. 
These rational solutions are special cases of the solutions of \PV\ \eqref{eq:p5} expressible in terms of the Kummer functions $M(a,b,z)$ and $U(a,b,z)$ discussed in \cite{refMasuda04,refOkamotoPV}.
\begin{definition}
The \textit{generalised Laguerre polynomial} $\Tmn{m,n}{\mu}(z)$, which is a polynomial of degree $(m+1)n$, is defined by 
\beq \label{def:Tmn}
\Tmn{m,n}{\mu}(z)=\det \left[\deriv[j+k]{}{z}L_{m+n}^{(\mu+1)} (z) \right]_{j,k=0}^{n-1}, \qquad m\ge -1, \quad n\geq 1,
\eeq
where $L_{n}^{(\mu)}(z)$ is the Laguerre polynomial %\eqref{eq:alp}.
\beq\label{eq:alp} \Lag{n}{\mu}(z)=\frac{z^{-\mu}\,\rme^z}{n!}\deriv[n]{}{z}\left(z^{n+\mu}\,\rme^{-z}\right),\qquad n\geq0.\eeq %see \cite[equation (13.6.19)]{refDLMF}
\end{definition}
%Properties of the generalised Laguerre polynomial $\Tmn{m,n}{\mu}(z)$ are discussed in \cite{refBB,refCD24}.

There are three classes of rational solutions of \PV\ \eqref{eq:p5} which are expressed in terms of generalised Laguerre polynomials:
\begin{subequations}\label{sols:GenLag}\begin{align}
{w}_{m,n}^{(\mu)}(z)&=\frac{T_{m-1,n}^{(\mu)}(z)\,T_{m-1,n+1}^{(\mu)}(z)}{\Tmn{m,n}{\mu}(z)\,T_{m-2,n+1}^{(\mu)}(z)},&&\pms{\hf m^2}{-\hf(m+2n+1+\mu)^2}{\mu},\\
v_{m,n}^{(\mu)}(z) &=\frac{m-n+\mu}{m+n+1}\,\frac{\Tmn{m,n}{\mu-2n}(z)\,\Tmn{m-1,n+1}{\mu-2n-2}(z)}{\Tmn{m-1,n}{\mu-2n}(z)\,\Tmn{m,n+1}{\mu-2n-2}(z)},&&\pms{\hf(m+n+1)^2}{-\hf(m-n+\mu)^2}{\mu},\\
 u_{m,n}^{(\mu)}(z) &=w_{m,n}^{(-\mu)}(-z), &&\pms{\hf m^2}{-\hf(m+2n+1-\mu)^2}{\mu}.
 \end{align}\end{subequations}
For further details see, for example, \cite{refCDM26}, which also gives alternative expressions for the rational solutions and the special cases when $m=0$ or $n=0$ are discussed.
 
The effect of applying the \bt\ $\Tr_{\ep_{1},\ep_{2},\ep_{3}}$ \eqref{eq:PT.BTs.P5} to the rational solutions $w_{m,n}^{(\mu)}(z)$ with $(a,b,c)=(m,m+2n+\mu+1,\mu)$, $v_{m,n}^{(\mu)}(z)$ with $(a,b,c)=(m+n+1,m-n+\mu,\mu)$ and $u_{m,n}^{(\mu)}(z)$ with $(a,b,c)=(m,m+2n-\mu+1,\mu)$ is given in Table \ref{BT_GenLag}. We note that the \bt\ $\Tr_{\ep_{1},\ep_{2},\ep_{3}}$ maps each rational solution to a rational solution in a different class. 
The effect of applying the Schlesinger transformations $\mathcal{R}_1,\mathcal{R}_{2},\ldots,\mathcal{R}_{14}$ to the rational solutions $w_{m,n}^{(\mu)}(z)$ with $(a,b,c)=(m,m+2n+\mu+1,\mu)$, $v_{m,n}^{(\mu)}(z)$ with $(a,b,c)=(m+n+1,m-n+\mu,\mu)$ and $u_{m,n}^{(\mu)}(z)$ with $(a,b,c)=(m,m+2n-\mu+1,\mu)$ is given in Table \ref{ST_GenLag}. 

\begin{table}[ht]
\[\begin{array}{|c|ccc|c|} \hline & \a_{1} & \b_{1} & \ga_{1} & {w}_{1} \\ \hline 
\Tr_{1,1,1}\big(w_{m,n}^{(\mu)}\big)& \hf (m+n)^2& -\hf(m+n+\mu)^2 & 2n+\mu+1 & v_{m-1,n}^{(\mu+2n+1)}\\
\Tr_{1,-1,1}\big(w_{m,n}^{(\mu)}\big)& \hf n^2&-\hf(n+\mu)^2 &2m+2n+\mu+1 & u_{n,m}^{(\mu+2m+2n+1)}\\
\Tr_{-1,1,-1}\big(w_{m,n}^{(\mu)}\big)& \hf(n+1)^2&-\hf(n+\mu+1)^2 &2m+2n+\mu+1 & u_{n+1,m-1}^{(\mu+2m+2n+1)}\\
\Tr_{-1,-1,-1}\big(w_{m,n}^{(\mu)}\big)& \hf (m+n+1)^2& -\hf(m+n+\mu+1)^2 & 2n+\mu+1 & v_{m,n}^{(\mu+2n+1)}\\ \hline
\Tr_{1,1,1}\big(v_{m,n}^{(\mu)}\big)&\hf m^2 &-\hf(m+\mu)^2 &-2n+\mu-1 &w_{m,n}^{(\mu-2n-1)}\\
\Tr_{1,-1,1}\big(v_{m,n}^{(\mu)}\big)&\hf(n+1)^2 &-\hf(n-\mu+1)^2 &2m+\mu+1 &u_{n+1,m}^{(\mu+2m+1)}\\
\Tr_{-1,1,-1}\big(v_{m,n}^{(\mu)}\big)& \hf n^2 &-\hf(n-\mu)^2 &2m+\mu+1 &u_{n,m}^{(\mu+2m+1)}\\
\Tr_{-1,-1,-1}\big(v_{m,n}^{(\mu)}\big)&\hf(m+1)^2 &-\hf(m+\mu+1)^2 & -2n+\mu-1 &w_{m+1,n}^{(\mu-2n-1)}\\ \hline
\Tr_{1,1,1}\big(u_{m,n}^{(\mu)}\big)& \hf(m+n-\mu)^2& -\hf(m+n)^2 & 2n-\mu+1&\wt{v}_{n,m-1}^{(\mu-2n-1)}\\
\Tr_{1,-1,1}\big(u_{m,n}^{(\mu)}\big)&\hf(n-\mu)^2 &-\hf n^2 & 2m+2n-\mu+1 &\wt{w}_{n,m}^{(\mu-2m-2n-1)}\\
\Tr_{-1,1,-1}\big(u_{m,n}^{(\mu)}\big)& \hf(n-\mu+1)^2 &-\hf (n+1)^2 & 2m+2n-\mu+1 &\wt{w}_{n+1,m-1}^{(\mu-2m-2n-1)}\\
\Tr_{-1,-1,-1}\big(u_{m,n}^{(\mu)}\big)&\hf(m+n-\mu+1)^2& -\hf(m+n+1)^2 & 2n-\mu+1 &\wt{v}_{n,m}^{(\mu-2n-1)}\\
\hline\end{array}\]
\caption{\label{BT_GenLag}The effect of applying the \bt\ $\Tr_{\ep_{1},\ep_{2},\ep_{3}}$ \eqref{eq:PT.BTs.P5} to the rational solutions $w_{m,n}^{(\mu)}(z;%)$ with $(a,b,c)=(
m,m+2n+\mu+1,\mu)$, $v_{m,n}^{(\mu)}(z;%)$ with $(a,b,c)=(
m+n+1,m-n+\mu,\mu)$ and $u_{m,n}^{(\mu)}(z;%)$ with $(a,b,c)=(
m,m+2n-\mu+1,\mu)$, where $\wt{w}_{n,m}^{(\mu)}(z)=1/{w}_{n,m}^{(\mu)}(z)$ and $\wt{v}_{n,m}^{(\mu)}(z)=1/{v}_{n,m}^{(\mu)}(z)$.}
\end{table}

\begin{table}[ht]
\[\begin{array}{|c|cccccccc|} \hline
& \mathcal{R}_{1} & \mathcal{R}_{2} & \mathcal{R}_{3} & \mathcal{R}_{4} & \mathcal{R}_{5} & \mathcal{R}_{6} & \mathcal{R}_{7} & \mathcal{R}_{8} \\ \hline
w_{m,n}^{(\mu)} & w_{m+1,n}^{(\mu-1)}&w_{m-1,n}^{(\mu+1)}&w_{m,n+1}^{(\mu-1)} &w_{m,n-1}^{(\mu+1)} & w_{m,n}^{(\mu-1)}&w_{m,n}^{(\mu+1)}&w_{m-1,n+1}^{(\mu-1)} &w_{m+1,n-1}^{(\mu+1)} \\
v_{m,n}^{(\mu)} & v_{m+1,n}^{(\mu-1)} & v_{m-1,n}^{(\mu+1)} & v_{m+1,n-1}^{(\mu-1)} & v_{m-1,n+1}^{(\mu+1)} & v_{m,n}^{(\mu-1)} & v_{m,n}^{(\mu+1)} & v_{m,n-1}^{(\mu-1)} & v_{m,n+1}^{(\mu+1)} \\
u_{m,n}^{(\mu)} & u_{m+1,n-1}^{(\mu-1)} & u_{m-1,n+1}^{(\mu+1)} & u_{m,n}^{(\mu-1)} & u_{m,n}^{(\mu+1)} & u_{m,n-1}^{(\mu-1)} & u_{m,n+1}^{(\mu+1)} & u_{m-1,n}^{(\mu-1)} & u_{m+1,n}^{(\mu+1)} \\
\hline
\end{array}\]
\[\begin{array}{|c|cccccc|} \hline
& \mathcal{R}_{9} & \mathcal{R}_{10} & \mathcal{R}_{11} & \mathcal{R}_{12} & \mathcal{R}_{13} & \mathcal{R}_{14} \\ \hline
w_{m,n}^{(\mu)} & w_{m,n+1}^{(\mu-2)}&w_{m,n-1}^{(\mu+2)}&w_{m+1,n}^{(\mu)} &w_{m-1,n}^{(\mu)} & w_{m+1,n-1}^{(\mu)}&w_{m-1,n+1}^{(\mu)}\\
v_{m,n}^{(\mu)} & v_{m+1,n-1}^{(\mu-2)} & v_{m-1,n+1}^{(\mu+2)} & v_{m+1,n}^{(\mu)} & v_{m-1,n}^{(\mu)} & v_{m,n+1}^{(\mu)} & v_{m,n-1}^{(\mu)} \\
u_{m,n}^{(\mu)} & u_{m,n-1}^{(\mu-2)} & u_{m,n+1}^{(\mu+2)} & u_{m+1,n}^{(\mu)} & u_{m-1,n}^{(\mu)} & u_{m+1,n-1}^{(\mu)} & u_{m-1,n+1}^{(\mu)} \\
\hline
\end{array}\]
\caption{\label{ST_GenLag}The effect of applying the Schlesinger transformations $\mathcal{R}_1,\mathcal{R}_{2},\ldots,\mathcal{R}_{14}$ to the rational solutions $w_{m,n}^{(\mu)}(z;%)$ with $(a,b,c)=(
m,m+2n+\mu+1,\mu)$, $v_{m,n}^{(\mu)}(z;%)$ with $(a,b,c)=(
m+n+1,m-n+\mu,\mu)$ and $u_{m,n}^{(\mu)}(z;%)$ with $(a,b,c)=(
m,m+2n-\mu+1,\mu)$.}
\end{table}

\subsection{\label{ssec:GenUm}Generalised Umemura polynomials}
Rational solutions in cases (ii) and (iii) of Theorem \ref{thm:KLM} are expressed in terms of \textit{generalised Umemura polynomials}.
Umemura \cite{refUm20} defined some polynomials through a differential-difference equation to describe rational solutions of \PV\ \eqref{eq:p5}; see also \cite{refPACpv,refNY98ii,refYamada}.
Subsequently these were generalised by Masuda, Ohta and Kajiwara \cite{refMOK}, who defined the {generalised Umemura polynomial} through coupled differential-difference equations and gave a representation as a determinant of Laguerre polynomials. Some properties of generalised Umemura polynomials are discussed in \cite{refCDM26,refMasuda25}.

Rather than use the generalised Umemura polynomial defined in \cite{refMOK} we use an equivalent representation given in Definition \ref{dfn:GenUm} below, which is obtained using matrix row and column operations. We find this representation simpler to compute and useful to prove algebraic and combinatorial properties of the polynomial, though we do not pursue this further here.

\begin{definition}{\label{dfn:GenUm}\rm The {\textit{generalised Umemura polynomial}} $\Umn{m,n}{\k}(z)$ is given by the Wronskian
\beq \Umn{m,n}{\k}(z)=c_{m,n}\exp(\hf mz)\Wr\!\left({\ph^{(\k)}_{1}},{\ph^{(\k)}_{3}},\ldots,{\ph^{(\k)}_{2m-1}};{\psi^{(\k)}_{1}},{\psi^{(\k)}_{3}},\ldots,{\psi^{(\k)}_{2n-1}}\right),\quad m,n\geq0, 
\label{def:Umn}\eeq
where 
\beq\label{def:cmn} c_{m,n}=(-1)^{m}\, 2^{(m+n)(m+n+1)/2}
\prod_{j=1}^{m-1}\frac{(2j+1)!}{j!}\prod_{k=1}^{n-1}\frac{(2k+1)!}{k!},\eeq
which makes the polynomial monic,
${\ph^{(\k)}_{n}(z)=\rme^{-z/2}\,L_{n}^{(\k)}(\hfz)}$ and ${\psi^{(\k)}_{n}(z)=L_{n}^{(\k)}(-\hfz)}$, 
with $\k$ a parameter and $\Lag{n}{\k}(z)$ the Laguerre polynomial \eqref{eq:alp} and $\Umn{0,0}{\k}(z)=1$.}
\end{definition}

There are three classes of rational solutions of \PV\ \eqref{eq:p5} which are expressed in terms of generalised Umemura polynomials
\begin{subequations}\label{sols:UmLag}\begin{align}
\wh{w}_{m,n}^{(\k)}(z)&=-\frac{\Umn{m,n-1}{2\k}(z)\,\Umn{m-1,n}{2\k+2}(z)}{\Umn{m,n-1}{2\k+2}(z)\,\Umn{m-1,n}{2\k}(z)},
&&\pms{\hf(m+\k)^2}{-\hf(n+\k)^2}{m+n},\\
\wh{v}_{m,n}^{(\k)}(z)&=-\frac{\Umn{m-1,n-1}{2\k-2n+3}(z)\,\Umn{m,n}{2\k-2n-1}(z)}{\Umn{m-1,n-1}{2\k-2n+1}(z)\,\Umn{m,n}{2\k-2n+1}(z)},
&&\pms{\hf(m+\k)^2}{-\hf(n-\k)^2}{m-n},\\
\wh{u}_{m,n}^{(\k)}(z) &=-\,\frac{\Umn{m,n-1}{\k+1}(z)\,\Umn{m,n+1}{\k-1}(z)}{\Umn{m-1,n}{\k+1}(z)\,\Umn{m+1,n}{\k-1}(z)},
&&\pms{\hf(m+\hf)^2}{-\hf(n+\hf)^2}{m+n+\k},
\end{align}\end{subequations} when $m\geq1$ and $n\geq1$.
These rational solutions were given in \cite{refMOK}; see also \cite{refCDM26}, where alternative expressions for the rational solutions are given and the special cases when $m=0$ or $n=0$ are discussed. Further we note that for $m\geq1$
\begin{subequations}\begin{align}
\wh{u}_{m,0}^{(\k)}(z)&=-\,\frac{\Umn{m,0}{\k}(z)\,\Umn{m,1}{\k-1}(z)}{\Umn{m-1,0}{\k+1}(z)\,\Umn{m+1,0}{\k-1}(z)}, &&\pms{\hf(m+\hf)^2}{-\tfrac{1}{8}}{m+\k},\label{sol:UmLag1} \\
\wh{u}_{0,0}^{(\k)}(z)&=-\,\frac{\Umn{0,1}{\k}(z)}{\Umn{1,0}{\k}(z)}=
-\frac{z+2\k+2}{z-2\k-2}, &&\pms{\tfrac{1}{8}}{-\tfrac{1}{8}}{\k+1}.\label{sol:UmLag0}
\end{align}\end{subequations}

The effect of applying the \bt\ $\Tr_{\ep_{1},\ep_{2},\ep_{3}}$ \eqref{eq:PT.BTs.P5} to the rational solutions $\wh{w}_{m,n}^{(\k)}(z)$ with $(a,b,c)=(m+\k,n+\k,m+n)$, $\wh{v}_{m,n}^{(\k)}(z)$ with $(a,b,c)=(m+\k,n-\k,m-n)$ and $\wh{u}_{m,n}^{(\k)}(z)$ with $(a,b,c)=(m+\hf,n+\hf,m+n+\k)$ is given in Table \ref{BT_GenUm}. As for rational solutions expressed in terms of generalised Laguerre polynomials, the \bt\ $\Tr_{\ep_{1},\ep_{2},\ep_{3}}$ maps each rational solution to a rational solution in a different class. The effect of applying the Schlesinger transformations $\mathcal{R}_1,\mathcal{R}_{2},\ldots,\mathcal{R}_{14}$ to the rational solutions $\wh{w}_{m,n}^{(\k)}(z)$ with $(a,b,c)=(m+\k,n+\k,m+n)$, $\wh{v}_{m,n}^{(\k)}(z)$ with $(a,b,c)=(m+\k,n-\k,m-n)$ and $\wh{u}_{m,n}^{(\k)}(z)$ with $(a,b,c)=(m+\hf,n+\hf,m+n+\k)$ is given in Table \ref{Sch_GenUm}. 

\begin{table}[ht]
\[\begin{array}{|c|ccc|c|} \hline & \a_{1} & \b_{1} & \ga_{1} & \wh{w}_{1} \\ \hline %(m+\k,n+\k,m+n)
\Tr_{1,1,1}\big(\wh{w}_{m,n}^{(\k)}\big) & \phantom{\Big(}\hf(\k-\hf)^2 \phantom{\Big(}& -\hf(m+n+\k-\hf)^2 &n-m& \wh{v}_{n,m}^{(-\k-n+1/2)} \\
\Tr_{1,-1,1}\big(\wh{w}_{m,n}^{(\k)}\big) & \hf(m+\hf)^2 & -\hf(n-\hf)^2 & m+n+2\k& \wh{u}_{m,n-1}^{(2\k+1)} \\
\Tr_{-1,1,-1}\big(\wh{w}_{m,n}^{(\k)}\big) & \hf(m-\hf)^2 & -\hf(n+\hf)^2 & m+n+2\k& \wh{u}_{m-1,n}^{(2\k+1)} \\
\Tr_{-1,-1,-1}\big(\wh{w}_{m,n}^{(\k)}\big) & \hf(\k+\hf)^2 & -\hf(m+n+\k+\hf)^2 &n-m& \wh{v}_{n,m}^{(-\k-n-1/2)} \\ \hline
\Tr_{1,1,1}\big(\wh{v}_{m,n}^{(\k)}\big) & \hf(n-\hf)^2& -\hf(m-\hf)^2 & n-m-2\k & \wh{u}_{n-1,m-1}^{(-2\k-2m+2)} \\
\Tr_{1,-1,1}\big(\wh{v}_{m,n}^{(\k)}\big) & \hf(m-n+\k+\hf)^2 & -\hf(\k+\hf)^2 & m+n & \wh{w}_{m,n}^{(\k-n+1/2)} \\
\Tr_{-1,1,-1}\big(\wh{v}_{m,n}^{(\k)}\big) & \hf(m-n+\k-\hf)^2 & -\hf(\k-\hf)^2 & m+n& \wh{w}_{m,n}^{(\k-n-1/2)} \\
\Tr_{-1,-1,-1}\big(\wh{v}_{m,n}^{(\k)}\big) & \hf(n+\hf)^2& -\hf(m+\hf)^2 & n-m-2\k& \wh{u}_{n,m}^{(-2\k-2m)}\\ \hline
\Tr_{1,1,1}\big(\wh{u}_{m,n}^{(\k)}\big) & \phantom{\Big(}\tfrac{1}{8}\k^2\phantom{\Big(} & -\tfrac{1}{8}(2m+2n+\k)^2 & n-m& \wh{v}_{n,m}^{(-\k/2-n)} \\
\Tr_{1,-1,1}\big(\wh{u}_{m,n}^{(\k)}\big) & \tfrac{1}{8}(2m+\k+1)^2 & -\tfrac{1}{8}(2n+\k-1)^2 & m+n+1& \wh{w}_{m+1,n}^{(\k/2-1/2)}\\
\Tr_{-1,1,-1}\big(\wh{u}_{m,n}^{(\k)}\big) & \tfrac{1}{8}(2m+\k-1)^2 & -\tfrac{1}{8}(2n+\k+1)^2 & m+n+1& \wh{w}_{m,n+1}^{(\k/2-1/2)}\\
\Tr_{-1,-1,-1}\big(\wh{u}_{m,n}^{(\k)}\big) & \tfrac{1}{8}(\k-2)^2 & -\tfrac{1}{8}(2m+2n+\k+2)^2 & n-m& \wh{v}_{n+1,m+1}^{(-\k/2-n)}\\
\hline\end{array}\]
\caption{\label{BT_GenUm} The effect of applying the \bt\ $\Tr_{\ep_{1},\ep_{2},\ep_{3}}$ \eqref{eq:PT.BTs.P5} to the rational solutions $\wh{w}_{m,n}^{(\k)}(z;%)$ with $(a,b,c)=(
m+\k,n+\k,m+n)$, $\wh{v}_{m,n}^{(\k)}(z;%)$ with $(a,b,c)=(
m+\k,n-\k,m-n)$ and $\wh{u}_{m,n}^{(\k)}(z;%)$ with $(a,b,c)=(
m+\hf,n+\hf,m+n+\k)$.}
\end{table}

\begin{table}[ht]
\[\begin{array}{|c|cccccccc|} \hline
& \mathcal{R}_{1} & \mathcal{R}_{2} & \mathcal{R}_{3} & \mathcal{R}_{4} & \mathcal{R}_{5} & \mathcal{R}_{6} & \mathcal{R}_{7} & \mathcal{R}_{8} \\ \hline
 \wh{w}_{m,n}^{(\k)} & \wh{w}_{m,n-1}^{(\k+1)}& \wh{w}_{m,n+1}^{(\k-1)}& \wh{w}_{m-1,n}^{(\k+1)} & \wh{w}_{m+1,n}^{(\k-1)} & \wh{w}_{m,n-1}^{(\k)}& \wh{w}_{m,n+1}^{(\k)}& \wh{w}_{m-1,n}^{(\k)} & \wh{w}_{m+1,n}^{(\k)} \\
 \wh{v}_{m,n}^{(\k)} & \wh{v}_{m,n+1}^{(\k+1)} & \wh{v}_{m,n-1}^{(\k-1)} & \wh{v}_{m,n+1}^{(\k)} & \wh{v}_{m,n-1}^{(\k)} & \wh{v}_{m-1,n}^{(\k+1)} & \wh{v}_{m+1,n}^{(\k-1)} & \wh{v}_{m-1,n}^{(\k)} & \wh{v}_{m+1,n}^{(\k)} \\
 \wh{u}_{m,n}^{(\k)} & \wh{u}_{m+1,n}^{(\k-2)} & \wh{u}_{m-1,n}^{(\k+2)} & \wh{u}_{m,n+1}^{(\k-2)} & \wh{u}_{m,n-1}^{(\k+2)} & \wh{u}_{m,n-1}^{(\k)} & \wh{u}_{m,n+1}^{(\k)} & \wh{u}_{m-1,n}^{(\k)} & \wh{u}_{m+1,n}^{(\k)} \\
\hline
\end{array}\]
\[\begin{array}{|c|cccccc|} \hline
& \mathcal{R}_{9} & \mathcal{R}_{10} & \mathcal{R}_{11} & \mathcal{R}_{12} & \mathcal{R}_{13} & \mathcal{R}_{14} \\ \hline
 \wh{w}_{m,n}^{(\k)} & \wh{w}_{m-1,n-1}^{(\k+1)}& \wh{w}_{m+1,n+1}^{(\k-1)}& \wh{w}_{m,n}^{(\k+1)} & \wh{w}_{m,n}^{(\k-1)} & \wh{w}_{m+1,n-1}^{(\k)}& \wh{w}_{m-1,n+1}^{(\k)}\\
 \wh{v}_{m,n}^{(\k)} & \wh{v}_{m-1,n+1}^{(\k+1)} & \wh{v}_{m+1,n-1}^{(\k-1)} & \wh{v}_{m+1,n+1}^{(\k)} & \wh{v}_{m-1,n-1}^{(\k)} & \wh{v}_{m,n}^{(\k+1)} & \wh{v}_{m,n}^{(\k-1)} \\
 \wh{u}_{m,n}^{(\k)} & \wh{u}_{m,n}^{(\k-2)} & \wh{u}_{m,n}^{(\k+2)} & \wh{u}_{m+1,n+1}^{(\k-2)} & \wh{u}_{m-1,n-1}^{(\k+2)} & \wh{u}_{m+1,n-1}^{(\k)} & \wh{u}_{m-1,n+1}^{(\k)} \\
\hline
\end{array}\]
\caption{\label{Sch_GenUm}The effect of applying the Schlesinger transformations $\mathcal{R}_1,\mathcal{R}_{2},\ldots,\mathcal{R}_{14}$ to the rational solutions $\wh{w}_{m,n}^{(\k)}(z;%)$ with $(a,b,c)=(
m+\k,n+\k,m+n)$, $\wh{v}_{m,n}^{(\k)}(z;%)$ with $(a,b,c)=(
m+\k,n-\k,m-n)$ and $\wh{u}_{m,n}^{(\k)}(z;%)$ with $(a,b,c)=(
m+\hf,n+\hf,m+n+\k)$.}
\end{table}

\section{\label{sec:Bessel_non-unique}Non-unique modified Bessel function solutions}
In this section we derive non-unique modified Bessel function solutions of \PV\ \eqref{eq:p5}, i.e.\ two distinct solutions in terms of modified Bessel functions, which are not related by the symmetries \eqref{PV:Sym}, that satisfy \PV\ for the same values of the parameters. Further we show that there is also a rational solution expressed in terms of generalised Umemura polynomials for these parameters.

Special function solutions of \PV\ \eqref{eq:p5} are usually expressed in terms of the 
Whittaker functions $M_{\kappa,\mu}(z)$ and $W_{\kappa,\mu}(z)$, or equivalently the Kummer functions $M(a,b,z)$ and $U(a,b,z)$, cf.~\cite[\S32.10(v)]{refDLMF}. 
However if $b=2a+n$, with $n\in\Integer$, then the Kummer functions $M(a,b,z)$ and $U(a,b,z)$ can be respectively expressed in terms of the modified Bessel functions $I_{\nu}(\hfz)$ and $K_{\nu}(\hfz)$, 
%for example %are special cases of Kummer functions 
%$$M(\nu+\tfrac12,2\nu+1,z)=\Gamma(1+\nu) (\tfrac14z)^{-\nu}I_{\nu}(\hfz)\exp(\hfz),\quad U(\nu+\tfrac12,2\nu+1,z)=\pi^{-1/2}z^{-\nu}K_{\nu}(\hfz)\exp(\hfz); $$ 
see \cite[\S13.6(iii)]{refDLMF}.
%\subsection{Relationship between Kummer and Bessel Functions}
Specifically, if $b=2a=2\nu+1$, then it is well known that
\begin{subequations}\begin{align} 
M\!\left(\nu+\hf,2\nu+1,2z\right)&=\Gamma\!\left(\nu+1\right)\rme^{z}(\hfz)^{-\nu}I_{\nu}(z),\\
U\!\left(\nu+\hf,2\nu+1,2z\right)&=\pi^{-1/2}\rme^{z}(2z)^{-\nu}K_{\nu}(z),
\end{align}\end{subequations}
with $I_{\nu}(z)$ and $K_{\nu}(z)$ modified Bessel functions. %, see \cite[equations 13.6.9, 13.6.10]{refDLMF}
More generally, if $a=\nu+\hf$ and $b=2a+n=2\nu+n+1$, with $n\in\mathbb{Z}$, then for $n\geq0$
\begin{subequations}\begin{align} 
M\!\left(\nu+\hf,2\nu+1+n,2z\right)&=\Gamma(\nu)\rme^{z}(\hfz)^{-\nu}\sum_{k=0}^{n}\sbr{-1}^k \binom{n}{k}\frac{{(2\nu)_{k}}(\nu+k)}{{(2\nu+1+n)_{k}}}I_{\nu+k}(z),\\
M\!\left(\nu+\hf,2\nu+1-n,2z\right)&=\Gamma(\nu-n)\rme^{z}(\hfz)^{n-\nu} %\\ &\qquad\times
\sum_{k=0}^{n} \binom{n}{k}\frac{{(2\nu-2n)_{k}}(\nu-n+k)}{{\left(2\nu+1-n\right)_{k}}}I_{\nu+k-n}(z),\\%[2.5pt]
U\!\left(\nu+\hf,2\nu+1+n,2z\right) &= \frac{2\,\poch{\nu+\hf }{n}\,\rme^z (2z)^{-\nu}}{\sqrt{\pi}}\sum_{k=0}^{n}\binom{n}{k}\frac{{\nu+k}}{\poch{2\nu+k}{n+1}}K_{\nu+k}(z),\\
U\!\left(\nu+\hf,2\nu+1-n,2z\right) &= \frac{2\,\sbr{-1}^n\,\rme^{z}(2z)^{n-\nu}}{\sqrt{\pi}} %\\&\qquad \times 
\sum_{k=0}^{n}\sbr{-1}^k\binom{n}{k}\frac{{\nu+k-n}}{\poch{2\nu+k-2n}{n+1}}K_{\nu+k-n}(z).
\end{align}\end{subequations}

It is convenient to define the function $\BesselZ{\nu}{z}$ when discussing solutions of \PV\ \eqref{eq:p5} which involve modified Bessel functions.
\begin{definition} Define the function $\BesselZ{\nu}{z}$ by
\beq\label{def:BesselZ}\BesselZ{\nu}{z}=
\begin{cases}C_{1} I_{\nu}(z)+C_{2}I_{-\nu}(z), & \text{if}\quad \nu\notin\Integer,\\
C_{1} I_{n}(z)+C_{2}(-1)^nK_{n}(z),\quad & \text{if}\quad n\in\Integer,\quad n\geq0,\end{cases}\eeq
where $I_{n}(z)$ and $K_{n}(z)$ are modified Bessel functions,
with $C_{1}$ and $C_{2}$ arbitrary constants.
\end{definition}

If in \PV\ \eqref{eq:p5}, $\a=-\b=\tfrac{1}{8}$, then $a=\pm\hf$ and $b=\pm\hf$, independently. Hence if we use an appropriate \bt\ of \PV\ \eqref{eq:p5} that changes the values of $a$ and/or $b$ by $\pm1$, then $a$ and/or $b$ can switch sign, so $\a$ and/or $\b$ are unchanged. This is illustrated in the following example. %\red{Change working since $\ga$ changes in the following example.}

\begin{example}\label{example:w0}
Consider the solution
\beq w_{0}(z)= \dfrac{\Zn{1}-\Zn{0}}{\Zn{1}+\Zn{0}},\label{sol:w0}\eeq
with $\BesselZ{\nu}{z}$ given by \eqref{def:BesselZ},
which satisfies \PV\ \eqref{eq:p5} for the parameters $\pms{\tfrac{1}{8}}{-\tfrac{1}{8}}{0}$. Applying the \bts\ $\Tr_{1,1,1}$ and $\Tr_{-1,-1,-1}$, with $(a,b,c)=(\hf,-\hf,0)$, which lead to the same values of $(\a,\b,\ga)$, to $w_{0}(z)$ gives
\begin{subequations}\label{sols:wvu1}\begin{align}
w_{1}(z)&=\Tr_{1,1,1}(w_{0};\hf,-\hf,0)=1-z+z\dfrac{\Zn{1}}{\Zn{0}},\\
v_{1}(z)&=\Tr_{-1,-1,-1}(w_{0};\hf,-\hf,0)=\dfrac{\Zn{0}}{z\Zn{1}+(z+1) \Zn{0}}.
\end{align}
These solutions both satisfy \PV\ \eqref{eq:p5} for the parameters $\pms{\tfrac{1}{8}}{-\tfrac{1}{8}}{-1}$ since
\[ \Tr_{1,1,1}(\hf,-\hf,0)=(\hf,-\hf,-1), \qquad \Tr_{-1,-1,-1}(\hf,-\hf,0)=(-\hf,\hf,-1).\]
We note that $ {v}_{1}(z)=1/{w_{1}(-z)}$, after redefining the constants, since for $n\in\Integer$
\[ I_{n}(-z)=(-1)^nI_{n}(z),\qquad K_{n}(-z)=(-1)^nK_{n}(z)-\rmi\pi I_{n}(z), \]
cf.~\cite[\S10.34]{refDLMF}, so the solutions are equivalent due to the symmetries \eqref{PV:Sym}.
The rational solution
\beq u_{1}(z)=\Tr_{1,1,1}(-1;\hf,-\hf,0)=-\frac{z-2}{z+2},\eeq\end{subequations}
which is not a special case of either $w_{1}(z)$ or $v_{1}(z)$,
also satisfies \PV\ \eqref{eq:p5} for the parameters $\pms{\tfrac{1}{8}}{-\tfrac{1}{8}}{-1}$. Further $u_{1}(z)$ is the special case of a rational solution expressed in terms of Umemura polynomials given by $u_{1}(z)=\wh{u}_{0,0}^{(-2)}(z)$, recall \eqref{sol:UmLag0}.

\comment{Next we apply two \textit{different} \bts\ to $w_{1}(z)$ which lead to two \textit{different} Bessel function solutions that satisfy \PV\ \eqref{eq:p5} for the \textit{same} parameters. The result is
\begin{subequations}\label{sols:wvu2}\begin{align}
w_{2}(z)&=\Tr_{1,1,1}^{\,3}(w_{1};\hf,\hf,-1)=-\frac{\Zn{1}-\Zn{0}}{\Zn{1}+\Zn{0}},\\
v_{2}(z)&=\Tr_{1,1,1}^{\,2}\circ\Tr_{-1,1,1}(w_{1};\hf,-\hf,-1)=\frac{\big[\Zn{1}+\Zn{0}\big] \big[z\Zn[2]{1}- (z-2)\Zn[2]{0}\big]}%
{\big[\Zn{1}-\Zn{0}\big]\big[z\Zn[2]{1}-(z+2)\Zn[2]{0} \big]},
\end{align}}%\end{subequations}

Next we apply the \bt\ $\Tr_{1,1,1}^{\,3}$, with $(a,b,c)=(\hf,\hf,-1)$, to the three solutions $w_{1}(z)$, $v_{1}(z)$ and $u_{1}(z)$ given by \eqref{sols:wvu1} and obtain the solutions
\begin{subequations}\label{sols:wvu2}\begin{align}
w_{2}(z)&=\Tr_{1,1,1}^{\,3}(w_{1};\hf,\hf,-1) =-\frac{\Zn{1}-\Zn{0}}{\Zn{1}+\Zn{0}},\label{sol:w2}\\
%=-1+\frac{2\Zn{0}}{\Zn{1}+\Zn{0}},\\
v_{2}(z)&=\Tr_{1,1,1}^{\,3}(v_{1};\hf,\hf,-1)=\frac{\big[\Zn{1}+\Zn{0}\big] \big[z\Zn[2]{1}- (z-2)\Zn[2]{0}\big]}{\big[\Zn{1}-\Zn{0}\big]\big[z\Zn[2]{1}-(z+2)\Zn[2]{0} \big]},\label{sol:v2}\\
%&=1-\frac{2\Zn{0}}{\Zn{1}-\Zn{0}}+\frac{4\Zn{0} \big[ z\Zn{1}+(z+1)\Zn{0} \big]}{z\Zn[2]{1}- (z-2)\Zn[2]{0}},\nonumber\\
u_{2}(z)&=\Tr_{1,1,1}^{\,3}(u_{1};\hf,\hf,-1)=-\frac{z-4}{z+4},
\end{align}\end{subequations}
which all satisfy \PV\ \eqref{eq:p5} for the parameters $\pms{\tfrac{1}{8}}{-\tfrac{1}{8}}{-2}$ since $\Tr_{1,1,1}^{\,3}(\hf,\hf,-1)=(\hf,-\hf,-2)$. 
Hence we have three solutions, two distinct solutions in terms of modified Bessel functions and a rational solution, which all satisfy \PV\ \eqref{eq:p5} for the same values of the parameters.

We remark that $w_{2}(z)=-w_{0}(z)$ and the solutions $w_{2}(z)$ and $v_{2}(z)$ given by \eqref{sol:w2} and \eqref{sol:v2} respectively are related by the \bt\ $\Tr_{-1,1,-1}\circ \Tr_{-1,1,1}$, with $(a,b,c)=(\hf,\hf,-2)$, since
\[\Tr_{-1,1,-1}\circ \Tr_{-1,1,1}(w_2;\hf,\hf,-2)=(v_2;-\hf,\hf,-2).\]
Further $u_{2}(z)=\wh{u}_{0,0}^{(-3)}(z)$, recall \eqref{sol:UmLag0}.

\end{example}
\begin{example}Applying the \bt\ $\Tr_{-1,1,-1}\circ\Tr_{-1,-1,1}$, with $(a,b,c)=(\hf,\hf,-2)$, to the three solutions $w_{2}(z)$, $v_{2}(z)$ and $u_{2}(z)$ given by \eqref{sols:wvu2} we obtain the solutions
\begin{subequations}\label{sols:wvu3}\begin{align}
w_{3}(z)&=-\frac{(2z-1)\Zn{1}-(2z-3)\Zn{0}}{3\big[\Zn{1}-\Zn{0}\big]}
=-\frac{2z-1}{3}-\frac{2\Zn{0}}{3\big[\Zn{1}-\Zn{0}\big]},
\label{sol:w3}\\
v_{3}(z) &=-\frac{\big[ (2z+1)\Zn{1}+(2z+3)\Zn{0}\big] \big[z\Zn[2]{1}-(z+2)\Zn[2]{0}\big]}{\big[\Zn{1}+\Zn{0}\big]\big[z(4z-1)\Zn[2]{1}+8z(z+1)\Zn{0}\Zn{1}+(4z^2+9z+6)\Zn[2]{0}\big]}\nonumber\\
&=-\frac{2z+1}{4z-1}+\frac{2\Zn{0}}{3\big[\Zn{1}+\Zn{0}\big]}\nonumber\\
&\qquad +\frac{\big[4z(2z+1)(2z+7)\Zn{1}+4(4z^3+20z^2+21z+3)\Zn{0}\big]\Zn{0}}{3(4z-1)\big[z(4z-1)\Zn[2]{1}+8z(z+1)\Zn{0}\Zn{1}+(4z^2+9z+6)\Zn[2]{0}\big]},\label{sol:v3}\\
u_{3}(z)&= -\frac{(z^2-12)(z+4)}{z^3+12z^2+36z+48} 
=-1+\frac{8(z^2+6z+12)}{z^3+12z^2+36z+48}, \label{sol:u3}
\end{align} \end{subequations}
respectively, which all satisfy \PV\ \eqref{eq:p5} for the parameters $\pms{\tfrac{9}{8}}{-\tfrac{1}{8}}{-2}$. 
Hence we have shown that for these parameters, \PV\ \eqref{eq:p5} has two different Bessel function solutions and a rational solution, which is expressed in terms of generalised Umemura polynomials since 
$u_{3}(z)=\wh{w}_{2,0}^{(-1/2)}(-z)=\wh{u}_{1,0}^{(-3)}(z)$,
recall \eqref{sol:UmLag1}.

The solutions $w_{3}(z)$, $v_{3}(z)$ and $u_{3}(z)$ given by \eqref{sols:wvu3} have different expansions about both $z=0$ and $z=\infty$. Setting $C_{2}=0$ in $w_{3}(z)$ and $v_{3}(z)$, since $K_{\nu}(z)$ has a singularity at $z=0$, then as $z\to0$
\begin{subequations}\begin{align} 
w_{3}(z)&=1-\frac{z}{2}+\frac{z^2}{24}+\frac{z^{3}}{192}-\frac{z^{5}}{9216}+\mathcal{O}\big(z^{6}\big),\\
v_{3}(z)&=1-\frac{z}{2}+\frac{z^2}{24}+\frac{19z^{3}}{192} -\frac{5z^{4}}{64}+\frac{95 z^{5}}{3072} +\mathcal{O}\big(z^{6}\big),\\
u_{3}(z)&=1-\frac{z}{2}+\frac{z^2}{24}+\frac{5z^{3}}{96}-\frac{5z^{4}}{128} +\frac{71 z^{5}}{4608}+\mathcal{O}\big(z^{6}\big),
\end{align}\end{subequations}
and as $z\to\infty$
\begin{subequations}\begin{align} 
w_{3}(z)&= -\frac{1}{2 z}-\frac{3}{2 z^{2}}-\frac{51}{8 z^{3}}-\frac{273}{8 z^{4}}-\frac{3483}{16 z^{5}}+\mathcal{O}\big(z^{-6}\big),\\
v_{3}(z)&=\frac{1}{2 z}+\frac{1}{2 z^{2}}+\frac{3}{8 z^{3}}+\frac{11}{8 z^{4}}+\frac{83}{16 z^{5}}+\mathcal{O}\big(z^{-6}\big),\\
u_{3}(z)&=-1+\frac{8}{z}-\frac{48}{z^{2}}+\frac{384}{z^{3}}-\frac{3264}{z^{4}}+\frac{27648}{z^{5}}+\mathcal{O}\big(z^{-6}\big).
\end{align}\end{subequations}
We note that the first three terms in the series expansions about $z=0$ are the same for all three solutions $w_{3}(z)$, $v_{3}(z)$ and $u_{3}(z)$ whereas as $z\to\infty$, $w_{3}(z)\to0$, $v_{3}(z)\to0$ and $u_{3}(z)\to-1$. 
A plot of the solutions $w_{3}(z)$, $v_{3}(z)$ and $u_{3}(z)$, with $C_{2}=0$ %(since $K_{\nu}(z)$ has a singularity at $z=0$), 
is given in Figure \ref{fig:wvu3}. All three solutions have the same value at $z=0$. The solution $w_{3}(z)$ has one zero for $z>0$ and a minimum at $z\approx 5.2855$. The solutions $v_{3}(z)$ and $u_{3}(z)$ are both monotonically decreasing for $z>0$.

\begin{figure}[ht]
\[\includegraphics[width=4in]{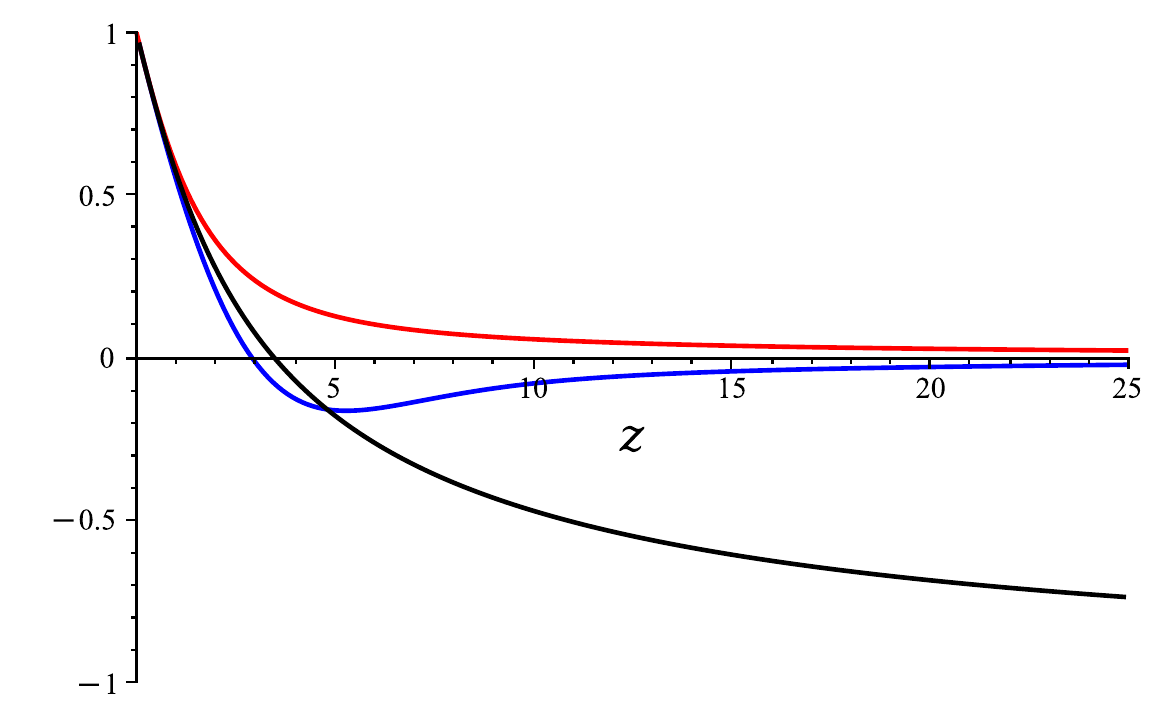}\]
\caption{\label{fig:wvu3}A plot of the solutions $w_{3}(z)$ (\blue{blue}), $v_{3}(z)$ (\red{red}) and $u_{3}(z)$ (black) given by \eqref{sols:wvu3}, with $C_{2}=0$ (since $K_{\nu}(z)$ has a singularity at $z=0$).}
\end{figure}
\end{example}

\begin{remarks}\label{rem43}
\begin{enumerate}\item[]
\item The rational solution $u_{3}(z)$ is not a special case of either $w_{3}(z)$ or $v_{3}(z)$ and analogously for the solutions $u_{1}(z)$ and $u_{2}(z)$ in Example \ref{example:w0}. Usually when a \peq\ has a special function solution and a rational solution for the same values of the parameters, then the rational solution is a special case of the special function solution. For example, all the rational solutions of \PV\ \eqref{eq:p5} in terms of generalised Laguerre polynomials given in \eqref{sols:GenLag}
are special cases of the Kummer function solutions of \PV\ \eqref{eq:p5}, see \cite{refCD24}.
\item The solution $w_{0}(z)$ given by \eqref{sol:w0} can be mapped to a solution of \PIII\ \eqref{eq:p3} as shown in Example \ref{exam54}.
\end{enumerate}
\end{remarks}

\section{\label{sec:PVPIII}Relationship with the third \peq}
In this section we discuss the relationship between solutions of the fifth \peq\ \eqref{eq:p5gen} and solutions of the generic case of \PIII\ \eqref{eq:p3} when $CD\not=0$.

It is well known that solutions of the deg-\PV\ \eqref{eq:p5deg}
%\beq\deriv[2]{u}{x}=\left(\frac{1}{2u} + \frac{1}{u-1}\right)\!\left(\deriv{u}{x}\right)^{\!\!2} -
%\frac{1}{x}\deriv{u}{x}\ + \frac{(u-1)^2(A u^2+B)}{x^2u} + \frac{C u}{x},\label{eq:degPV}\eeq
%which is equation \eqref{eq:p5gen} with $\de=0$, 
are related to solutions of \PIII\ \eqref{eq:p3}
with $CD\not=0$, a result originally due to Gromak \cite{refGromak75}; see also \cite[\S34]{refGLS}. 
This relationship is described in the following lemma.

\begin{lemma}{\label{lem21}If $q(x)$ is a solution of %\eqref{eq:p31} 
 \beq\deriv[2]{q}{x} = \frac{1}{q}\! \left(\deriv{q}{x}\right)^{\!2}- \frac{1}{x} \deriv{q}{x}
 +\frac{(a-b)q^2}{x} + \frac{a+b+1}{x} + \frac{q^3}{4} - \frac{1}{4q},\label{eq:p31}\eeq
which is \PIII\ \eqref{eq:p3} with $(A,B,C,D)=(a-b,a+b+1,\tfrac{1}{4},-\tfrac{1}{4})$,
%$\a=(a-b)$, $\b=(a+b+1)$, $\ga=\tfrac{1}{4}$ and $\de=-\tfrac{1}{4}$,
then
\beq\label{tr:qu} u(z)=%\frac{zq'+ \hfz q^2+(2a+2b+1)q- \hfz}{zq'- \hfz q^2+(2a+2b+1)q- \hfz}
1+\frac{2xq^2}{\ds 2x\deriv{q}{x}-xq^2+2(2a+2b+1)q-x},\qquad z=\tfrac{1}{4}x^2,\eeq
%with $'\equiv \rmd/\rmd z$, 
is a solution of %\eqref{eq:degPV1}, 
 \beq\deriv[2]{u}{z}= \left(\frac{1}{2u} + \frac{1}{u-1}\right)\!\left(\deriv{u}{z}\right)^2 - \frac{1}{z}\,\deriv{u}{z} + 
\frac{(u-1)^2(a^2 u^2 -b^2)}{2z^2u}+\frac{u}{2z},\label{eq:degPV1}\eeq
which is {\rm deg}-\PV\ \eqref{eq:p5deg} with $\a=\tfrac12a^2$, $\b=-\tfrac12b^2$ and $\ga=\hf$, provided that 
\[ x\deriv{q}{x}- \hf x q^2+(2a+2b+1)q- \hf x\not=0.\]
Conversely, if $u(z)$ is a solution of \eqref{eq:degPV1}, then
\beq\label{tr:uq} q(x)=\frac{1}{\sqrt{z}\,u}\left\{z\deriv{u}{z} +(u-1)(au+b)\right\} ,
%=\frac{\sqrt{x}}{u}\,\deriv{u}{x} + \frac{(u-1)(au+b)}{\sqrt{x}\,u},
\qquad x=2\sqrt{z},\eeq
is a solution of \eqref{eq:p31}.
 }\end{lemma}
 
 \begin{proof} See, for example, \cite[Lemma 2.1]{refPAC23}, \cite[Theorem 4.2]{refFA82}, \cite[Theorem 34.2]{refGLS}.
 \end{proof}
 
 \begin{remark}We remark that \eqref{tr:qu} and \eqref{tr:uq} can be written as the system
\begin{subequations}\begin{align}
x\deriv{q}{x}&=\frac{\sqrt{z}(u+1)q^2}{u-1}-(2a+2b+1)q+\sqrt{z},\\
z\deriv{u}{z}&=-au^2+(\hf xq+a-b)u+b.
\end{align}\end{subequations}
Eliminating $u$ gives \eqref{eq:p31}, whilst eliminating $q$ gives \eqref{eq:degPV1}.
 \end{remark}

\begin{lemma}\label{lem:dPV}
There is a relationship between solutions of \PV\ \eqref{eq:p5} with $\a=-\b=\la$ and $\ga=0$, and solutions of {\rm deg}-\PV\ \eqref{eq:p5deg} with $\a=0$, $\b=-4\la$ and $\ga=\hf$.
\end{lemma}

\begin{proof}
Consider \PV\ \eqref{eq:p5} with $\a=-\b=\la$ and $\ga=0$, i.e.
\beq\wzz=\left(\frac{1}{2w}+ \frac{1}{w-1}\right)\!\left(\wz\right)^{\!\!2}-
\frac{1}{z}\wz+ \frac{\la(w-1)^3(w+1)}{z^2w} -\frac{w(w+1)}{2(w-1)}.\label{eq:p51}\eeq
Making the transformation
\beq u(x)=\frac{4w(z)}{[1+w(z)]^2},\qquad x=\tfrac{1}{4}z^2, \label{tr:wu}\eeq
in deg-\PV\ \eqref{eq:p5deg} with $\a=0$, $\b=-4\la$ and $\ga=\hf$ gives \eqref{eq:p51}.
%\[\wzz=\left(\frac{1}{2w}+ \frac{1}{w-1}\right)\!\left(\wz\right)^{\!\!2}-\frac{1}{z}\wz- \frac{B(w-1)^3(w+1)}{4z^2w} -\frac{2w(w+1)}{w-1},\] 
%which is \eqref{eq:p51}. 
Conversely, making the transformation
\[ w(z)=\frac{1-\sqrt{1-u(x)}}{1+\sqrt{1-u(x)}},\qquad z=2\sqrt{x},\]
in \eqref{eq:p51} gives
\[\deriv[2]{u}{x}=\left(\frac{1}{2u} + \frac{1}{u-1}\right)\!\left(\deriv{u}{x}\right)^{\!\!2} - \frac{1}{x}\deriv{u}{x} -\frac{4\la(u-1)^2}{x^2 u} +\frac{u}{2x},\]
which is deg-\PV\ \eqref{eq:p5deg} with $\a=0$, $\b=-4\la$ and $\ga=\hf$.
\end{proof}

From Lemmas \ref{lem21} and \ref{lem:dPV} we have the following corollary.
\begin{corollary}
There is a relationship between solutions of \PV\ \eqref{eq:p5} with $\a=-\b$ and $\ga=0$ and solutions of \PIII\ \eqref{eq:p3} with $A+B=1$ and $C=-D=\tfrac{1}{4}$.
\end{corollary}

\begin{proof}
If $w(z)$ is a solution of \eqref{eq:p51} with $\la=\tfrac{1}{8}k^2$, then substituting \eqref{tr:wu} into \eqref{tr:uq} with $a=0$ and $b=k$ gives
\beq q(x)=-\frac{w(z)-1}{w(z)[w(z)+1]}\wz -\frac{k[w(z)-1]^2}{2zw(z)},\qquad x=z,\label{tr:wq}\eeq
which satisfies \PIII\ \eqref{eq:p3} with $(A,B,C,D)=(-k,k+1,\tfrac{1}{4},-\tfrac{1}{4})$. 

To derive the inverse transformation, from \eqref{tr:qu} and \eqref{tr:wu}
\[ \frac{4w(z)}{[1+w(z)]^2}=
1+\frac{2xq^2(x)}{\ds 2x\deriv{q}{x}-xq^2(x)+2(2k+1)q(x)-x},\]
we obtain
\beq w(z)=\frac{\sqrt{Q(x)}+1}{\sqrt{Q(x)}-1},\qquad Q(x)=-\frac{1}{q^2(x)}\deriv{q}{x}+\frac{q^2(x)+1}{2q^2(x)}-\frac{2k+1}{xq(x)},\qquad z=x.\label{tr:qw}\eeq
Therefore if $q(x)$ satisfies \PIII\ \eqref{eq:p3} with $(A,B,C,D)=(-k,k+1,\tfrac{1}{4},-\tfrac{1}{4})$, then $w(z)$ given by \eqref{tr:qw} satisfies \eqref{eq:p51} with $\la=\tfrac{1}{8}k^2$, which is \PV\ \eqref{eq:p5} with $\pms{\tfrac{1}{8}k^2}{-\tfrac{1}{8}k^2}{0}$.
\end{proof}

Using ideas due to Tsegel'nik \cite{refTsegel25}, there is another case when solutions of \PV\ \eqref{eq:p5} with $\ga=0$ are related to solutions of \PIII\ \eqref{eq:p3}.
\begin{lemma} Suppose that $q(x)$ is a solution of \PIII\ \eqref{eq:p3} with $(A,B,C,D)=(\hf\la,-\hf\la,\tfrac{1}{4},-\tfrac{1}{4})$, where $\la$ is an arbitrary constant, then
\beq\label{lem55w} w(z)=1+\frac{2[q^2(x)-1]}{\ds 2\deriv{q}{x}-q^2(x)+1},\qquad z=2x,\eeq
satisfies \PV\ \eqref{eq:p5} with $(\a,\b,\ga)=\big(\tfrac{1}{8}(\la-1)^2,-\tfrac{1}{8}(\la+1)^2,0\big)$. The inverse transformation is given by
\beq\label{lem55q} \frac{q^2(x)+1}{4q(x)}= \frac{1}{2w(z)}\deriv{w}{z}+\frac{(\la-1)[w(z)-1]}{4z}-\frac{(\la+1)[w(z)-1]}{4zw(z)},\qquad x=\hf z.\eeq
\end{lemma}

\begin{proof} It is straightforward to show that if $w(z)$ is given by \eqref{lem55w}, with $q(x)$ a solution of \PIII\ \eqref{eq:p3} with $(A,B,C,D)=(\hf\la,-\hf\la,\tfrac{1}{4},-\tfrac{1}{4})$, then it satisfies
\beq\label{eq:55w} \wzz=\left(\frac{1}{2w}+\frac{1}{w-1}\right)\!\left(\wz\right)^{\!\!2}-\frac{1}{z}\wz+\frac{(w-1)^2\big\{(\la-1)^2w^2- (\la+1)^2\big\}}{8z^2w} - \frac{w(w+1)}{2(w-1)},\eeq
which is \PV\ \eqref{eq:p5} with $(\a,\b,\ga)=\big(\tfrac{1}{8}(\la-1)^2,-\tfrac{1}{8}(\la+1)^2,0\big)$. 

To prove the inverse transformation \eqref{lem55q}, consider 
\[ v(z)=\frac{1}{2w(z)}\deriv{w}{z}+\frac{(\la-1)[w(z)-1]}{4z}-\frac{(\la+1)[w(z)-1]}{4zw(z)},\]
where $w(z)$ satisfies \eqref{eq:55w}, then it is straightforward to show that $v(z)$ satisfies
\beq\label{eq:55v} \deriv[2]{v}{z}=\frac{4v}{4v^2-1}\left(\deriv{v}{z}\right)^{\!2} -\frac{1}{z}\deriv{v}{z}+\frac{(zv+\la)(4v^2-1)}{4z}.\eeq
Consequently making the transformation
\[v(z)=\frac{q^2(x)+1}{4q(x)},\qquad z=2x,\]
in equation \eqref{eq:55v} gives
\beq\label{eq:55q} \deriv[2]{q}{x}=\frac{1}{q}\!\left(\deriv{q}{x}\right)^{\!2}-\frac{1}{x}\deriv{q}{x}+\frac{\la (q^2-1)}{2x}+\frac{q^3}{4}-\frac{1}{4q},\eeq
which is \PIII\ \eqref{eq:p3} with $(A,B,C,D)=(\hf\la,-\hf\la,\tfrac{1}{4},-\tfrac{1}{4})$. 
\end{proof}

\begin{remarks}\textrm{
\begin{enumerate}
\item We remark that making the transformation
\[ v(z)=\frac{w(z)+1}{2[w(z)-1]},\]
in equation \eqref{eq:55v} gives
\[ \wzz=\left(\frac{1}{2w}+\frac{1}{w-1}\right)\!\left(\wz\right)^{\!\!2}-\frac{1}{z}\wz- \frac{\la w}{z} - \frac{w(w+1)}{2(w-1)},\]
which is \PV\ \eqref{eq:p5} with $(\a,\b,\ga)=\big(0,0,-\la\big)$. 
%\end{remark}\begin{remark}
\item \p\ \cite[pp.\ 39--40]{refPain02} showed that \PV\ \eqref{eq:p5} with $\a=\b=0$ and $\ga=c$ can be transformed into \PIII\ \eqref{eq:p3} through a quadratic transformation.
Substituting 
\[ w(z)= \left\{\frac{q(x)+1}{q(x)-1}\right\}^{\!2},\qquad z=2x,\]
into \PV\ \eqref{eq:p5} with $\pms{0}{0}{c}$ %\a=\b=0$ and $\ga=c$, 
gives equation \eqref{eq:55q} with $\la=-c$.
%\[ \deriv[2]{q}{z}=\frac{1}{q}\!\left(\deriv{q}{z}\right)^{\!2}-\frac{1}{z}\deriv{q}{z}-\frac{c(q^2-1)}{4z}+\frac{q^3}{16}-\frac{1}{16q},\]
%which is \PIII\ \eqref{eq:p3} with $(A,B,C,D)=(-\tfrac{1}{4} c,\tfrac{1}{4} c,\tfrac{1}{16},-\tfrac{1}{16})$.
\end{enumerate}}\end{remarks}

\begin{example}\label{exam54} In Example \ref{example:w0} we saw that the solution
\[ w_{0}(z)= \frac{\Zn{1}-\Zn{0}}{\Zn{1}+\Zn{0}},\]
with $\BesselZ{\nu}{z}$ given by \eqref{def:BesselZ},
satisfies \PV\ \eqref{eq:p5} for the parameters $\pms{\tfrac{1}{8}}{-\tfrac{1}{8}}{0}$. Hence we can use \eqref{tr:wq}, with $\k=\pm1$, to derive a solution of \PIII\ \eqref{eq:p3}. Substituting $w_{0}(z)$ into \eqref{tr:wq} with $\k=-1$ gives
\[ q(x)= -\frac{\Zx{0}}{\Zx{1}},\]
which satisfies \PIII\ \eqref{eq:p3} with $(A,B,C,D)=(1,0,\tfrac{1}{4},-\tfrac{1}{4})$, whilst substituting $w_{0}(z)$ into \eqref{tr:wq} with $\k=1$ gives
\[ q(x) %= -\frac{\Zx{0}\big\{z\Zx[2]{1}+4\Zx{1}\Zx{0}-z\Zx[2]{0} \big\}}{\Zx{1}\big\{\Zx[2]{1}-\Zx[2]{0} \big\}}
= -\frac{\Zx{0}}{\Zx{1}}-\frac{4\Zx[2]{0}}{x\big\{\Zx[2]{1}-\Zx[2]{0} \big\}} ,\]
which satisfies \PIII\ \eqref{eq:p3} with $(A,B,C,D)=(-1,2,\tfrac{1}{4},-\tfrac{1}{4})$.
\end{example}

\begin{example}If $w_0$ is given by \eqref{sol:w0}, then consider the function
\[ w(z) =\Tr_{1,-1,1}^2(w_0;\hf,-\hf,0)
=- \frac{\big\{\Zn{1}+\Zn{0}\big\}\big\{(z-2)\Zn[2]{1}+4\Zn{1}\Zn{0}-z\Zn[2]{0}\big\}}{\big\{\Zn{1}-\Zn{0}\big\}\big\{(z+2)\Zn[2]{1}+4\Zn{1}\Zn{0}-z\Zn[2]{0} \big\}},\]
with $\BesselZ{\nu}{z}$ given by \eqref{def:BesselZ},
satisfies \PV\ \eqref{eq:p5} for the parameters $\pms{\tfrac{9}{8}}{-\tfrac{9}{8}}{0}$. Substituting this into \eqref{tr:wq} with $\k=3$ gives
\begin{align*} q(x) &= \frac{\Zx{1}\big\{ (x^2+8)\Zx[2]{1}+2x\Zx{1}\Zx{0}-x^2\Zx[2]{0}\big\}}{x\big\{2\Zx[3]{1}-x\Zx[2]{1}\Zx{0}-4\Zx{1}\Zx[2]{0}+x\Zx[3]{0} \big\} }\\&\qquad -
\frac{8\Zx[2]{1}\left\{(x^2+2)\Zx[2]{1}+2x\Zx{1}\Zx{0}-x^2\Zx[2]{0}\right\}}{x\left\{4\Zx[4]{1}-\left[x\Zx[2]{1}+4\Zx{1}\Zx{0}-x\Zx[2]{0}\right]^2 \right\}},
\end{align*}
which satisfies \PIII\ \eqref{eq:p3} with $(A,B,C,D)=(-3,4,\tfrac{1}{4},-\tfrac{1}{4})$, and substituting it into \eqref{tr:wq} with $\k=-3$ gives
\begin{align*} q(x) = \frac{4\Zx[2]{1}}{x\big\{\Zx[2]{0}-\Zx[2]{1}\big\}}
+\frac{\Zx{1}\big\{ (x^2+8)\Zx[2]{1}+2x\Zx{1}\Zx{0}-x^2\Zx[2]{0}\big\}}{x\big\{2\Zx[3]{1}-x\Zx[2]{1}\Zx{0}-4\Zx{1}\Zx[2]{0}+x\Zx[3]{0} \big\} },
\end{align*}
which satisfies \PIII\ \eqref{eq:p3} with $(A,B,C,D)=(3,-2,\tfrac{1}{4},-\tfrac{1}{4})$.
\end{example}

\begin{theorem}{\rm\label{thm:GLS415}
\textit{The equation \PV\ \eqref{eq:p5} reduces to a special case of \PIII\ \eqref{eq:p3} when the parameters $\a$, $\b$, $\ga$ and $\de$ of \PV\ take one of the following forms:}
\begin{enumerate}[(i)]
\item \textit{if $\sqrt{-2\b}=\sqrt{2\a} + m$, $\ga = n$ and $\de = -\hf$, where $m,n\in\mathbb{Z}$ and $m+n$ is even;}
\item \textit{if $\a=\tfrac{1}{8}(2m-1)^2$, $\b=-\tfrac{1}{8}(2n-1)^2$, $\ga\notin\mathbb{Z}$ and $\de=-\hf$, where $m,n\in\mathbb{Z}$.}
\end{enumerate}
}\end{theorem}

\begin{proof} See \cite[Theorem 41.5]{refGLS}. \end{proof}

\begin{remarks}\textrm{
\begin{enumerate}
\item Whilst the conditions in Theorem \ref{thm:GLS415}(i) are sufficient, from the comments in Remarks \ref{rem:2.3}, when either $\sqrt{2\a}=\pm\hf$ or $\sqrt{-2\b}=\pm\hf$ there are other cases for which the condition that $m+n$ is even is not necessary.
\item The parameters in Theorem \ref{thm:GLS415}(ii) are the same as for case (iii) in the classification of rational solutions of \PV\ \eqref{eq:p5} given in Theorem \ref{thm:KLM} which are expressed in terms of generalised Umemura polynomials. 
The proof in \cite[Theorem 41.5]{refGLS} says that all these rational solutions have the seed solution $w=-1$ with $\a=-\b$ and $\ga=0$. %, which contradicts the condition that $\ga\notin\mathbb{Z}$. 
This is also a rational solution of \PIII\ \eqref{eq:p3} if $A=-B$ and $C=-D$, so the link between the two sets of solutions is rather tenuous.
\end{enumerate}}\end{remarks}

\section{\label{sec:dis}Discussion}
In this paper we studied the \bts\ of the fifth Painlev\'e equation \PV\ \eqref{eq:p5} defining their action on the parameters $a=\sqrt{2\a}$, $b=\sqrt{-2\b}$ and $c=\ga$, rather than on $\a$, $\b$ and $\ga$. 
This removes the ambiguity in the choice of square roots and makes it straightforward to compose transformations. Using compositions of the eight elementary transformations $\mathcal{T}_{\ep_{1},\ep_{2},\ep_{3}}$ we gave explicitly the combinations which shift two of the parameters $a$, $b$ and $c$ by $\pm1$.
We also identified the Schlesinger transformations $\mathcal{R}_{1},\mathcal{R}_{2},\ldots,\mathcal{R}_{14}$ obtained from the isomonodromy problem, and the translation operators of the symmetric form \spv\ \eqref{eq:spv} associated with the extended affine Weyl group $\widetilde{W}\big(A_{3}^{(1)}\big)$, with specific compositions of these \bts. 
This gives a unified description of the \bts\ of \PV\ \eqref{eq:p5} that appear in the literature in several different notations.
As an application, we applied these \bk\ and Schlesinger transformations to the rational solutions of \PV\ \eqref{eq:p5}, which are expressed in terms of generalised Laguerre polynomials and generalised Umemura polynomials. 

An interesting observation in the paper concerns the case when $\a=-\b=\tfrac18$, so that $a=\pm\tfrac12$ and $b=\pm\tfrac12$ independently. In this case a \bt\ can change the sign of $a$ and/or $b$ without changing $\a$ and $\b$. Consequently two different transformations applied to the same seed solution can produce distinct solutions for the same values of the parameters. 
In this way we obtained for the same parameters, two distinct solutions expressed in terms of the modified Bessel functions $I_{0}(z)$, $I_{1}(z)$, $K_{0}(z)$ and $K_{1}(z)$. Further, for the same parameters, we showed that there is also a rational solution, expressible in terms of generalised Umemura polynomials, which is not a special case of either Bessel function solution. 
This contrasts with the usual situation for \peqs, when there is both a special function solution and a rational solution for the same parameters, then the rational solution is a special case of the special function solution.

Finally, we discussed the relationship between \PV\ \eqref{eq:p5} and \PIII\ \eqref{eq:p3} with $CD\neq0$. Using the known relationship between \PIII\ and deg-\PV\ \eqref{eq:p5deg}, we clarified a direct relationship between \PV\ \eqref{eq:p5} with $\a=-\b$, $\ga=0$ and \PIII\ \eqref{eq:p3} with $A+B=1$, $C=-D=\tfrac14$. Motivated by the work of Tsegel'nik \cite{refTsegel25}, we obtained a second relationship for \PV\ \eqref{eq:p5} with $\pms{\tfrac18(\lambda-1)^2}{-\tfrac18(\lambda+1)^2}{0}$.

\subsection*{Declarations}
\textbf{Acknowledgements.}
BM is grateful to EPSRC for the support of Doctoral Training Partnerships grant EP/W524050/1.

\smallskip\noindent\textbf{Data availability statement.}
No data were used for the research described in the article.

\smallskip\noindent\textbf{Declaration of competing interest.}
The authors declare no conflicts of interest.

\subsection*{ORCID iDs} 

\begin{tabular}{ll}
Peter A.\ Clarkson & 0000-0002-8777-5284 \\
Clare Dunning & 0000-0003-0535-9891
\end{tabular}

%\appendix

\def\refjl#1#2#3#4#5#6#7{\vspace{-0.2cm}
\bibitem{#1}\textrm{\frenchspacing#2}, \textrm{#3},
\textit{\frenchspacing#4}, \textbf{#5}\ (#6) #7.}

\def\refbk#1#2#3#4#5{\vspace{-0.2cm}
\bibitem{#1} \textrm{\frenchspacing#2}, ``\textit{#3}", #4 (#5).}

\def\refcf#1#2#3#4#5#6#7{\vspace{-0.2cm}
\bibitem{#1} \textrm{\frenchspacing#2}, \textrm{#3},
in ``\textit{#4}", {\frenchspacing#5}, #6\ (#7).}

\def\refpp#1#2#3#4{\vspace{-0.2cm}
\bibitem{#1} \textrm{\frenchspacing#2}, \textrm{#3}, #4.}

\def\refirnm#1#2#3#4#5{\vspace{-0.2cm}
\bibitem{#1}\textrm{\frenchspacing#2}, \textrm{#3},
\textit{\frenchspacing Int. Math. Res. Not. IMRN} (#4)\ #5.}

\def\refjsp#1#2#3#4#5{\vspace{-0.2cm}
\bibitem{#1}\textrm{\frenchspacing#2}, \textrm{#3},
\textit{\frenchspacing J. Stat. Mech.} (#4)\ #5.}

\def\refjsm#1#2#3#4#5#6#7{\vspace{-0.2cm}
\bibitem{#1}\textrm{\frenchspacing#2}, \textrm{#3},
\textit{\frenchspacing#4} (#6)\ #7.}

\def\refjlta#1#2#3#4#5#6#7{\vspace{-0.2cm}
\bibitem{#1}\textrm{\frenchspacing#2}, \textrm{#3},
\textit{\frenchspacing#4}, accepted for publication.} %to appear.} %\textbf{#5}\ (#6) #7.}

\def\fit#1{\textit{\frenchspacing#1}}

{}

\end{document}